# Single-shot, coherent, pop-out 3D metrology

Deepan Balakrishnan[1,2*], See Wee Chee[1,3], Zhaslan Baraissov[1,3], Michel Bosman[4], Utkur Mirsaidov[1,2,3,4], and N. Duane Loh[1,2,3*]

[1]Centre for Bio-Imaging Sciences, National University of Singapore, 117557, Singapore
[2]Department of Biological Sciences, National University of Singapore, 117557, Singapore
[3]Department of Physics, National University of Singapore, 117551, Singapore
[4]Department of Materials Science and Engineering, National University of Singapore, 117575, Singapore

[*]Corresponding authors. Email: *deepan@u.nus.edu*, *duaneloh@nus.edu.sg*

## Abstract

Three-dimensional (3D) imaging of thin, extended specimens at nanometer resolution is critical for applications in biology, materials science, advanced synthesis, and manufacturing. One route to 3D imaging is tomography, which requires a tilt series of a local region. However, capturing images at higher tilt angles is infeasible for such thin, extended specimens. Here, we explore a suitable alternative to reconstruct the 3D volume using a single, energy-filtered, bright-field coherent image. We show that when our specimen is homogeneous and amorphous, simultaneously inferring local depth and thickness for 3D imaging is possible in the near-field limit. We demonstrated this technique with a transmission electron microscope to fill a glaring gap for rapid, accessible 3D nanometrology. This technique is applicable, in general, to any coherent bright field imaging with electrons, photons, or any other wavelike particles.

## Introduction

Reconstructing three-dimensional (3D) information about a sample from two-dimensional (2D) projections is an important class of problems in optics. Though 3D imaging finds applications in various fields, from medicine to robotics, rapid large-scale 3D reconstruction is especially critical for nanoscience and nanotechnology. Since the dimensions of a nano-object define its properties and functionalities, rapid structural feedback is important for designing fabrication and synthesis strategies, failure analyses, reverse-engineering, and counterfeit verification [1–3]. Further, large-scale 3D nanometrology is useful for investigating nanometer-scale features on millimetre-scale objects (e.g., nanophotonic structures on butterfly wing scales and integrated circuit components) [4–7]. However, existing nanometrology tools cannot satisfy the demands of the semiconductor industry as it moves toward *3D power scaling* to make denser and more energy-efficient compute and storage components [8,9].

One route to 3D reconstruction is tomography, which is made possible due to the emergence of computing. Over the years, tomographic techniques improved along with computing power as they are heavily reliant on computation [10–13]. Consequently, today, X-ray computed tomography and electron tomography are widely used tools for 3D structural analysis at the nanometer scale, [14–18]. However, the time required to acquire a full-tilt series typically limits the resolvable dynamics to a few seconds, which is insufficient for imaging deformation dynamics that require millisecond resolution [19,20]. More important for extended samples common in the fabrication of electronic components, where there is usually a high aspect ratio between at least one pair of dimensions, collecting a full tilt-series is challenging either because of occlusion by the sample holder at high tilt angles or X-ray/electron absorption when viewing down the extended dimension. Ptychographic X-ray laminography, a coherent imaging technique, still requires a series of tilt images[4] and can only be performed at large bright X-ray facilities such as synchrotrons and X-ray free-electron lasers[21].

With machine learning, researchers have demonstrated that using priors as constraints can produce 3D tomography even with a vastly reduced number of angular views [11]. This points to the possibility of achieving 3D metrology with even a single view when there are sufficiently strong priors. Holography, which already encodes the depth information of a specimen, provides a viable direction for this form of single-shot 3D metrology.

Gabor showed in 1948 that shining a coherent wave, be it light or electron, through a specimen produces a single interference pattern which encodes the depth information in the recorded phases [22]. However, only after the advent of digital image capture and computational reconstruction did the holographic principle become widely exploited for various imaging applications [23,24]. Similarly, the multislice beam propagation formalism for modelling multiple scattering, well-known since 1957, has recently emerged as a viable tool for dealing with thicker and denser samples [25–28]. Again only after the phenomenal rise of computing power, more sophisticated algorithms and modelling schemes for digital holography were developed, where the 3D positions of thousands of particles (of known morphology) were inferred to nanometer-scale precision at millisecond rates [29–31].

Although these lensless holographic schemes allow a specimen's depth information to be readily inferred, its thickness information is far less apparent. Without both depth and thickness information, 3D metrology of a general specimen from a single view is infeasible. Compressive holography techniques enabled 3D reconstruction from a single 2D hologram for sparse 3D densities [32,33]. However, for densely packed specimens, decompressive interference using a forward model is an ill-posed problem [34]. On the contrary, estimating the local depth information and the optical thickness of the scatterer from a single projection, rather than reconstructing its phase shifts, is feasible. Here, we show that simultaneously inferring local depth and thickness for 3D metrology is possible in the near-field limit, which can be attained by having an objective lens in an in-line holography setup. This setup is readily described by a transmission electron microscope (TEM); with field emission guns generating highly coherent electron beams, TEMs effectively produce magnified in-line near-field holograms [35]. Further, direct electron detectors, because of their linear response and high detective quantum efficiency at all spatial frequencies, produce high-quality images [36,37].

We present a coherent imaging technique that allows rapid 3D metrology of amorphous, thin, extended materials from a single bright field image. We experimentally demonstrated the technique on thicker (200 nm) fabricated specimens to achieve a 3D resolution of 30 nm. For specimens thinner than 40nm, sub-10 nm resolution is possible at an imaging dose of 100 e Å$^{-2}$. We coin this technique *single-shot pop-out 3D metrology*, which simultaneously measures both the thickness and the imaging depth (i.e., *z*-position) of the material by utilising both the absorption and phase contrast information from a TEM image formed by a partially coherent beam. While the local thickness of a specimen is routinely measured from its amplitude contrast in Bright Field -TEM (BF-TEM)[38], estimating its local depth along the optical axis is far less common. Because this method extracts 3D information from a single 2D image, the same field of view can be re-interrogated rapidly, which allows us to study fast structural dynamics of materials with nanometer resolution. The data efficiency of pop-out metrology also supports fast *in-situ* inline process control and defect detection in nanofabrication of 3D devices over extended surfaces without physically sectioning or fracturing the sample as one would in critical dimension metrology. Finally, this technique can be readily calibrated for the many TEMs that are commonly found in many manufacturing, fabrication, and research facilities.

## Results and Discussion

### Principle of pop-out 3D metrology

Here, we briefly show how this depth information is encoded within the coherent interference pattern that is the TEM image, similar to that in inline digital holographic microscopy. Consider the kinematic linearization (the first Born approximation) of the dynamical multislice scattering of a sample comprising a $z$-stack of $N$ slices of random, amorphous scatterers[25,26](details in methods). We arrive at Eq. (1) for the squared amplitude of the Fourier transform of its TEM image

$$|I_{\text{det}}(\boldsymbol{k})|^2 \propto \mathrm{B}(\boldsymbol{k}) + \Gamma(\boldsymbol{k}) \equiv \sum_{n=1}^{N} \beta_n(\boldsymbol{k}) + \sum_{n=1}^{N}\sum_{m>n}^{N} \gamma_{mn}(\boldsymbol{k}) . \tag{1}$$

Eq. (1) can be decomposed into two key sums, each with an intuitive interpretation. The dominant sum $\mathrm{B}(\boldsymbol{k})$ adds the single scattering of the electron wavefield from each slice, $\beta_n(\boldsymbol{k})$, which arrives at the detector after various path differences. Crucially, we show that $\mathrm{B}(\boldsymbol{k})$ dominates the effective contrast transfer function (CTF) in our model (Eq. (2)), where the CTF's effective defocus is offset by the half-depth of the stack of slices. Fitting for this offset is key to inferring the depth of the sample.

$$|I_{\text{det}}(\boldsymbol{k})| \approx A \ \mathrm{env}(k) \ |\mathrm{CTF}(\boldsymbol{k})| + \mathrm{noise}(k), \tag{2}$$

Next, we turn our attention to the double sum $\Gamma(\boldsymbol{k})$ in Eq. (1), which captures the wavefield interference between different pairs of slices $\gamma_{mn}(\boldsymbol{k})$. It turns out that $\Gamma(\boldsymbol{k})$ tends to average away for multiple, random, uncorrelated slices (Fig. S1, Supplementary Note 1). Further, if the slices have comparable scattering densities, then the summation in Eq. (1) can thus be simplified (see methods), and what remains is an effective CTF in Eq. (2), which also includes the effects of lens aberrations and beam incoherence (see methods). Note that the visibility of these Thon rings is strongly affected by the spatial coherence of the electron source in TEMs[39]. Moreover, it is important to note that the accurate estimation of CTF in Eq. (2) is challenging with detectors with nonlinear response curves (e.g., scintillator-based detectors). The extension of the pop-out principle to multiple layers is shown in Eq. (40).

With the sample's local depth information inferred, we now infer the sample's local thickness. This thickness can be calibrated from the fraction of electrons lost to inelastic scattering (see methods), a procedure identical to inferring matter concentrations from optical absorbance via the Beer-Lambert law. To estimate this fraction with BF-TEM, we energy-filter for only elastically scattered electrons via hardware. The energy filtering enables us to estimate the sample thickness quantitatively.

Putting both the above depth and thickness inference together gives us the pop-out principle, which we illustrate in Fig. 1a using an amorphous specimen of homogeneous density. The specimen's right region is thinner and closer to the back focal plane than the left region. Hence, the thinner right region elastically scatters fewer electrons, which makes the right half of the energy-filtered TEM image appear brighter. Quantitatively, the relative thickness of these two regions is determined from the log-ratio of the number of electrons received at each region [38].

The depth of the centre of mass of each region in Fig. 1a is apparent in their local power spectrum (i.e., squared-amplitude of their Fourier transforms). As predicted by the CTF in Eq. (2), the Thon rings [40] in the power spectrum from the right region are spaced farther apart compared to those from the left region. This ring-spacing is related to the relative defocus of each region with respect to the TEM's plane of zero defocus. Consequently, each region's centre of mass depth can be determined from its defocus parameter from a semi-empirical fit to these Thon rings [13,41–44].

Combining the inferred local sample thickness and its centre of mass depth in Fig. 1a, we can correctly *pop-out* a thicker block of material on the left half of the field of view that is farther away than the thinner, nearer block on the right. Notice that with only thickness information alone, it is impossible to correctly place these two blocks at the correct depth. Repeating this recipe across small, overlapping patches in the 2D TEM image allows us to recover a 3D structure of the entire field of view[45]: the depths of these overlapping patches are stitched together to reconstruct how the specimen's centre of mass depth (i.e., along *z*-axis) varies transversely (i.e., along the *x-y* plane).

A proof of concept for pop-out 3D metrology is shown in Fig. 1b using a simulated BF-TEM image. Starting from a ground truth 3D object made of amorphous silicon nitride (bottom left of Fig. 1b), we used a multislice approach that includes absorption effects[39] to simulate its energy-filtered BF-TEM image. From this BF-TEM image, we inferred the patch-wise defocus and thickness maps of the 3D object, which were combined to *pop-out* the 3D reconstruction of the object (bottom right of Fig. 1b). Note that the structure of Fig. 1b cannot be retrieved from thickness information alone. The astute reader might have realized that although we illustrated the pop-out metrology principle here for the BF-TEM, the formalism of Eq. (1) applies generally whenever the multislice (i.e., beam propagation) description is apt and thus can be customized to the calibrated CTF of such optical setups beyond electron microscopy.

It is instructive to compare pop-out 3D metrology with phase retrieval and depth sectioning used for 3D imaging of nanomaterials[46–49]. The main difference is that these methods require a through-focal series for 3D reconstruction , while pop-out 3D metrology uses only a single TEM image.

## Factors affecting the resolution of pop-out metrology

Understanding which factors limit the resolution for a given specimen on a given microscope can help us calibrate its key parameters for optimal pop-out reconstructions. The in-plane ($xy$) and out-of-plane ($z$) resolutions of a pop-out reconstruction depend on the sample material, sample thickness, range of defocus parameters, patch size, and total electron dose chosen in the experiment. Here, we use realistic simulations to study the interplay between these factors and how they affect both in-plane and out-of-plane resolutions.

A critical factor in a pop-out reconstruction is the patch size. Since we determine the sample's average thickness and centre of mass depth over an image patch, the size of this patch limits our $xy$ resolution. Furthermore, the size of the patch is proportional to the total number of elastically scattered electrons within the patch, as well as how finely the Thon rings are sampled in the Fourier domain. Hence, the patch size, in turn, impacts the $z$ resolution of a pop-out reconstruction: the precision of each patch's determined centre of mass depth.

**Effect of sample thickness on the resolution**

The thickness of a specimen along the optical axis, $T_{\max}$, determines both the first node of the envelope function (dotted vertical lines in Fig. 2a) and the number of electron counts received at the detector. The envelope node suppresses the undulations in the Thon rings, which diminishes our ability to resolve the specimen's local depth ($z$) from the rings. Similarly, when the received electron counts are too low for model fitting, we must increase the patch size, which lowers our $xy$ resolution.

The effects on $xy$ and $z$ resolutions do not change linearly with sample thickness. Given a fixed incident electron dose, a thicker sample elastically scatters a larger fraction of the incident electrons, which imprints clearer Thon rings over the unscattered beam in the sample's TEM power spectrum. This is a positive effect because clear rings allow better fits for the sample's centre of mass depth. However, a thicker sample also inelastically scatters a larger fraction of the incident electrons, which reduces the number of electrons that reach the downstream image detector. This has a negative impact because the entire power spectrum is now noisier, which increases the uncertainty in our semi-empirical fits for centre of mass depths. Fig. 2c shows this positive effect dominates over the negative effect at modest sample thicknesses.

**Effect of spatial sampling and patch size on the resolution**

Increasing the image patch size increases the sampling frequency of the power spectrum, which improves their fit to the semi-empirical CTF functions (Fig. 2b). This, in turn, improves $z$ resolution but notably at the expense of $xy$ resolution. For a decent fit, we adopt a simple criterion for sampling frequency of at least three intensity maxima within the first node of the envelope (exemplified by Fig. 2a), and at least eight frequency samples between maxima ($\sigma_s = 8$) (see methods). To accommodate for unforeseen uncertainties in actual experiments, we recommend having slightly higher sampling frequencies (hence, larger image patch sizes).

In Fig. 2b, we performed a numerical experiment to analyse the influence of the image patch size on pop-out reconstructions. Here, we simulated multislice TEM images of an amorphous silicon nitride pillar (50 nm tall) on top of two different silicon nitride substrates (thicknesses of 25 nm or 100 nm). The corresponding maximum sample thicknesses are $T_{\max}$= 75 nm and 150 nm (pillar on the substrate), respectively, which lead to minimum image patch sizes of $w_{\min}$= 18 nm and 27 nm (see Eq. (39)). Fig. 2b illustrates the effect on pop-out reconstructions of the $T_{\max}$= 75 nm sample with sufficiently large patch sizes (20 nm patch in case i), or with sizes that are too small (12.8 nm in case ii). When the maximum sample thickness increases to $T_{\max}$= 150 nm (case iii), even the larger image patch of 20 nm produces a poorer reconstruction as it is less than the minimum patch size required (27 nm) for $T_{\max}$= 150 nm.

**Effect of electron dose on depth resolution**

The precision and accuracy of depth and thickness determination are impacted by the total number of elastically scattered electrons measured at the detector for each image patch. This number, in turn, depends on the total incident electron dose and the sample's thickness.

To study the effects of electron dose, we simulated a series of multislice TEM images with different specimen thicknesses and integrated electron dose. Fig. 2c shows the depth errors in our pop-out reconstructions of these images with an image patch size of 20 nm. This figure illustrates that 5 nm accuracy in $z$ depth is possible for silicon features that have a thickness less than half of the inelastic mean free path (in this case, 66.5 nm for SiNx) with exit doses measured at the detector that are ~100 e Å$^{-2}$. The error map plotted against exit dose instead of incident dose to account for electrons lost to inelastic scattering. The inelastic mean free path assumed for multislice simulations is 133 nm, which is calibrated from our experiments [3,50]. We expected a monotonous increase in the error values with further thickness increase. However, the vertical artefacts that show deviation from the expected trend highlight the limitation of our model. The precision in the error map (i.e., the standard deviation of the error calculated over a small range, say 5x5 data points, across thickness and dose in Fig. 2c) improves as the exit dose increases dramatically up to 100 e Å$^{-2}$, with negligible improvement with further dose increases. Finally, when comparing the reconstructions in Fig. 2b, it is evident that increasing the patch size lowers depth errors. However, this lowered error, which improves the $z$ resolution because of the larger patch size, is at the expense of $xy$ resolution. Although the studies of $xy$ and $z$ resolutions vs. electron dose and patch size in Fig. 2 were from idealized simulations, similar studies can be calibrated for known samples at actual TEMs. Such calibrations were performed for the applications shown below.

Pop-out demonstration

The process flow chart in Fig. 3 describes the steps and checks to implement pop-out metrology. Once the TEM is properly calibrated for capturing images, then the data acquisition (i.e., scanning a large area for the image capture, stitching the images, highlighted by the dashed red box in Fig. 3) and the pop-out reconstruction process can be fully automated. The pertinent details are elaborated as a checklist in the methods section.

**Dual-layer pop-out metrology**

The demonstrations of the pop-out principle have so far been limited to single-layered specimens. However, Eq. (40) shows that this principle can be readily extended to samples comprising multiple layers. This extension is corroborated in Fig. 4 for a two-layered sample made from a single type of amorphous material. We first consider a scheme for estimating the depth of each layer. Notably, the power spectrum of image patches within this sample exhibits intensity modulations that resembled those from the two layers separately (Fig. 4a). By modeling this power spectrum as an incoherent addition of two single-layered CTFs (see methods), we found that we can separately estimate the depth of each of the two layers. A full pop-out reconstruction, however, also requires the thickness information from both layers, despite measuring only the total fraction of electrons lost by both layers. Nevertheless, should the thickness of one of the layers be known (Fig. 4b) or the scattering from both layers be sufficient to fit the relative thickness between them (Fig. 4c), the thicknesses of the layers can be deduced.

Fig. 4b shows a proof of concept for a dual-layer pop-out with such a thickness prior. This sample comprises a 25-nm-thick top silicon nitride membrane (thickness known), and a lower membrane whose (unknown) thickness linearly increases in one direction from 2–50 nm. Our dual-layer pop-out principle correctly reconstructs the entire structure. The radial average of the fitted CTF from the additive model and the actual Thon rings from the wedge and membrane are plotted in Fig. 4a, which closely matches the incoherent additive modeling approach for multi-layered structures.

Fig. 4c shows a proof of concept for estimating the relative thickness of each of the two layers from the amplitudes of their respective CTFs in the additive-CTF fitting. This sample comprises two 20-nm-thick silicon nitride membranes with 40-nm-thick fins such that the total thickness in the projection is uniformly 80 nm across the specimen. Again, we note that the structure in Fig. 4c cannot be retrieved from thickness information alone. Since the amount of material across the layers are comparable, and we have enough signal from a large patch size and a high electron dose, our dual-layer pop-out principle correctly reconstructs the structure using the information from the ratio of the amplitudes without any thickness priors. Nevertheless, dual-layer pop-out is evidently more challenging compared to single-layer pop-out (Fig. S7, Supplementary Note 6).

**Experimental demonstrations**

We validated the pop-out principle in two proof-of-concept experiments. In each case, we sought to recover 3D features that are challenging to infer directly from their TEM images, other scanning probe measurements (i.e., AFM, SEM, etc.), or tomography (because of their high aspect ratio). The TEM images were collected with either a brief 2–6 s exposure or multiple brief exposures without having to rotate the sample. Taken together, these demonstrations justify why we coined our method *single-shot pop-out 3D metrology.*

First, we started with a relatively simple 3D nanochannel that was etched onto one side of an amorphous silicon nitride (a-SiN$_x$) substrate[3]. The leftmost panel of Fig. 5a shows a single-exposure TEM image of this nanochannel from which we reconstructed the 3D profile using the pop-out principle (the middle and rightmost panels of Fig. 5a show the top and bottom of the channel). The linear size of the floating cube

(reconstructed voxel size) in this reconstruction corresponds to the size of the image patch used for pop-out reconstruction, hence limiting the transverse $xy$ half-period resolution to 30 nm. Our reconstruction correctly shows that the specimen's bottom surface is flat, and that the nanochannel was etched from the top surface. This flat bottom side cannot be inferred from thickness information alone.  Incidentally, our reconstruction also shows that the substrate on which the nanochannel was etched has a small tilt (~5°).

In our second proof-of-principle demonstration, we wanted to reconstruct more complex 3D features than the first demonstration, again from a single-exposure TEM image. Here, we imaged a ~80 nm pit that was etched on an amorphous silicon nitride substrate. A TEM image of this pit and the corresponding pop-out reconstruction are shown in Fig. 5b, (resolution 40 nm, limited by patch size used in pop-out). The dark blob in the TEM image represents the debris from the etching piled near the rim of the pit, and the irregular etching around the rim causes the lighter petals. The cross-section shows that both the debris and the petals are on the top surface where the ion beam used for etching was incident. Our reconstruction reveals that the hollow region of the pit forms a double-conical structure: first narrowing as we etch deeper into the pit, then blowing-out to a larger width on the bottom side. Supplementary Note 5 shows evidence that this double-conical structure is not an artefact of pop-out reconstruction (Fig. S6). Moreover, this double-conical shape has also been reported in nanopore etching on a silicon nitride membrane [51]. This hidden feature, which is neither visible from the top surface nor easy to scan with a probe, is hard to measure using an AFM or SEM.

Combining these proof-of-principle experiments, we expect single-shot pop-out metrology to be able to rapidly recover the 3D structure of low-dimensional amorphous materials without rotating the sample and the usual sample preparation needed for tomography (e.g., ion milling, microtoming)[52]. These rapid 3D reconstructions can resolve nanometer longitudinal and transverse strain dynamics of micron-sized laminae that are stressed *in operando*, which complements atomic-resolution studies of the same but is restricted to a small region [53]. In principle, we can image millisecond 3D dynamics using pop-out metrology on a TEM with a kilohertz detector. However, as shown above, pop-out is only robust with sufficient signal per exposed frame. Speculating further, under the correct conditions and samples, it may even be possible to implement pop-out metrology with ultrafast electron microscopy (UEM), which currently achieves femtosecond 4DEM by repeatedly interrogating a rotating sample with an ultrafast electron pulse [20,53].

Although Pop-out metrology is a viable tool for 3D imaging of amorphous homogenous-density samples, unlike tomography, it cannot reconstruct the complete 3D densities of the specimen. Tomography and pop-out are complementary techniques, but the latter is not a replacement for the former. Pop-out metrology can be a viable alternative to tomography when we have specimens with homogeneous material density or for extended specimens where the missing-cone problem in tomography is severe (e.g., thin and extended samples). However, for general 3D samples where pop-out is under-constrained, tomography should still be the de facto 3D imaging option.

## Conclusions

Pop-out metrology is a computational wave optics technique that is applicable to coherent electron, neutron, X-ray, or visible light beams. At its core, it shows that the Fourier transform of an image formed by a coherent bright field illumination is essentially a hologram, and by exploiting the priors from the homogenously amorphous specimen, the depth information can be extracted even from a multi-layered specimen. In addition, the local thickness information can be measured in real space when there is an objective lens to focus the local absorption contrast on the image plane. Adding an objective lens with a known aberration model and a colour filter to remove the inelastically scattered photons allows pop-out metrology to be implemented in an inline digital holographic microscopy (DHM) setup, which further extends DHM's real-time live imaging capabilities[27].

We have experimentally demonstrated that the 3D density distribution of an amorphous sample that is a few hundred nanometers thick can be recovered to 30 nm resolution using only a single energy-filtered TEM image. This resolution increases for thinner samples, where sub-10 nm resolution is possible for samples thinner than 40nm. By automating this process, the field of view can be extended to several micrometers. Using the pop-out principle, 3D sample reconstruction is possible without having to rotate the sample (e.g., tomography or laminography) or destroy it (e.g., critical dimension metrology). We detailed this pop-out principle, the key imaging parameters that control resolution, and described how to generalize to multi-layer structures.

Considering how TEMs are already routinely used to characterize nanostructures in biology, material science, and semiconductor fabrication, we speculate that with suitable automation, this pop-out principle can be useful for fast 3D characterisation of the structural dynamics within a large field of view. The rapid feedback afforded by this pop-out technique with little to no sample modification on many existing TEMs makes it suitable as a fast screening tool, which fills an important gap amongst existing nanometer-scale metrology modalities. Furthermore, we speculate this method to be relevant for imaging nanometer features of complex structures commonly found in physical and biological sciences.

## Methods

The simulations are carried out using a TEM simulator that was developed in the programming language *Python*. All the simulations are generated for a microscope with energy 200 keV, spherical aberration of 1.2 mm, and a detector with pixel size of 6 μm. The experiments are carried out on a JEOL 2200 TEM equipped with a DE16 direct electron detector and an omega energy filter with a 20 eV window around the zero-loss peak. The 3D reconstructions are visualized using the TomViz application [54].

### Checklist

To optimize the resolution achieved from the pop-out metrology, the TEM parameters should be calibrated for the specimen. A checklist is provided here for this calibration, and a schematic of the process flow is shown in Fig. 3.

1. Ensure that the spherical aberration parameter $C_s$ of the TEM is known (see Eq. (1-3)). Otherwise, conduct experiments to fit the spherical aberration value for the TEM.
2. Ensure that the pixel size of the detector is known.
3. Calculate the resolution limit for the specimen thickness from Eq. (35) and determine the feasible range of magnification/resolution for the specimen given this resolution limit.
4. Calculate the theoretical reconstruction voxel size limit for the specimen from Eq. (39).
5. Ensure the dose limit for the specimen is known, and fix the total dose exposure for the experiment. We can now calculate the window size for the defocus fit, which should be just large enough to capture the required signal.
6. Before imaging the specimen, capture the electron beam without any specimen at the chosen magnification values for different electron doses. The electron beam might vary for various reasons in a TEM, and the electron dose cannot be measured accurately at every pixel due to different gain responses of the detector pixels. This series of images will help remove all the uncertainties if used instead of the total electron dose values.
7. Ensure that the energy filter is applied and choose the proper cutoff threshold for the specimen so that all the inelastically scattered electrons are filtered out.

8. While capturing the TEM images with the specimen, ensure that at least three prominent CTF rings are visible at every part of the TEM image by adjusting the defocus value.
9. Once we are set with the magnification, total dosage, and defocus value, we start capturing TEM images. For large specimens, capture a series of TEM images by lateral scanning. These TEM images can be stitched together and used as a single image during the reconstruction.
10. If the thickness of any part of the specimen is known, we can use that to calculate the electron mean free path in the material for the current TEM setting. Otherwise, we have to image a calibration specimen made of the same material with the known thickness.

The items in the checklist (except item 9) are all calibration steps, either routine (1,2,6) or sample-specific (3,4,5,7,8,10). We should clarify that steps 3 and 8 require us to know the maximum distance between the top and bottom of the sample; step 10 requires some thickness prior on the sample.

## Estimating sample depth from the defocus parameter in the contrast transfer function

Our method needs to determine the representative centre of the mass plane of scattering volume elements in each patch of pixels of the specimen in its TEM image. As shown below (in the section derivation of semi-empirical multislice scattering), the depth of this centre of scattering mass plane along the optical *z*-axis is encoded in spatial frequency domain $\boldsymbol{k} = (k_x, k_y)$ as the relative defocus $\Delta f$ parameter of the patch's Thon rings.

Assuming that a monochromatic electron plane wave of wavelength $\lambda$ impinges the sample, the spatial frequency $\boldsymbol{k}$-dependent contrast transfer function of its exit wave can be modeled by the semi-empirical model (Eq. (2)) in its azimuthally averaged form ($k = |\boldsymbol{k}|$)[55]. The CTF, envelope, and noise terms of the model are defined by

$$\mathrm{CTF}(\boldsymbol{k}) = w_1 \sin(\overline{\chi}) - w_2 \cos(\overline{\chi}), \quad \overline{\chi} = \pi\lambda k^2(0.5\lambda^2 k^2 C_S - \overline{\Delta f}) + \Delta\varphi, \tag{3}$$

$$\mathrm{env}(k) = e^{-(b_1 k + b_2 k^2 + b_3 k^4)}, \tag{4}$$

$$\mathrm{noise}(k) = n_4 e^{-n_1 \sqrt{k}} + n_5 e^{-n_2 k} + n_6 e^{-n_3 k^2}. \tag{5}$$

In Eq. (2), $A$ is an overall amplitude of elastically scattered electrons; the env term models the envelope caused by the relative thickness of the sample and the spatial, temporal coherence of the electron wave. The noise term models the cross multiplication term $\Gamma(\boldsymbol{k})$ (Eq. (29)) along with the background noise unrelated to the specimen; CTF is the contrast transfer function, $\chi$ is the aberration function that depends on spherical aberration $C_s$, defocus $\overline{\Delta f}$ ($\overline{\Delta f} = \Delta f + T/2$), specimen thickness $T$, electron wavelength $\lambda$ and any known overall phase shifts $\Delta\varphi$ (e.g., which might be caused by a phase plate). The values for the constants $w_1$ and $w_2$ are obtained from the amplitude contrast ratio $Q$, which is related to the ratio between the real and imaginary parts of the scattering potential $\epsilon$ (see the derivation section).

To account for astigmatism, the defocus $\Delta f$ term in the aberration function can be modified as:

$$\overline{\Delta f} \to \left[\overline{\Delta f_0} + \overline{\Delta f_{\mathrm{ast}}}\cos(2[\Theta_k - \Theta_{\mathrm{ast}}])\right], \tag{6}$$

where $\overline{\Delta f_0}$ and $\overline{\Delta f_{\mathrm{ast}}}$ are the average defocus value and effective astigmatism which is half of the difference between defocuses in major and minor axes, $\Theta_{\mathrm{ast}}$ is the angle between the major axis and the *x*-axis, and $\Theta_k$ is the angle between the scattering vector and *x*-axis [13,41–43]. We compare power spectrums from a simulated and an experiment micrograph with their corresponding fits which accounted for astigmatism in Fig. S2 (Supplementary Note 2).

Overall, from our derivation (shown below), we see that $\overline{\Delta f_0}$ in Eq. (6) is the relative defocus, or depth, of the sample's centre of scattering mass from the focal plane ($\overline{\Delta f_0} = \Delta f_0 + T/2$). From Fig. 2a and Fig. S1, we can see that the attenuation of CTF by a sinc-like envelope suppresses the Thon rings, which in turn reduces the signal for depth estimation. The derivation for this sinc-like envelope's node position (Eq. (30)) enabled us to deduce the sampling criterion and the resolution limit for thicker specimens (Eq. (39)). In Supplementary Note 3 (Fig. S3), we show that this sinc-like function can be approximated with a Gaussian envelope (Eq. (4)).

Notably, the cross multiplication term $\Gamma(\mathbf{k})$ is negligible for amorphous specimens since there is no correlation between the scattering potentials of different slices. This assumption, however, is not true for general crystalline materials. Hence, an increase in the crystallinity of the material affects the performance of the defocus estimation negatively, as experimentally demonstrated by Supplementary Note 4.2 (Fig. S5).

## Thickness estimation

The thickness map of the specimen is determined from the absorption contrast i.e., from the ratio of elastically scattered electrons.

$$T_{(x,y)} = \ell_{\mathrm{mfp}} \ln(I_0(x,y)/I_\mathrm{t}(x,y)), \tag{7}$$

where $I_0(x,y)$ is the total electron dose, $I_\mathrm{t}(x,y)$ is the transmitted unabsorbed electrons detected in a BF-TEM image, and $\ell_{\mathrm{mfp}}$ is the inelastic electron mean free path of the specimen. In Eq. (7), it is assumed that all the inelastically scattered electrons are removed before reaching the detector. Inelastically scattered electrons form an overall background that does not directly contribute depth or thickness information about the sample. Hence, an energy filter should be applied to filter the inelastically scattered electrons; otherwise, an error will be introduced in the thickness map (Supplementary Note 4). We compare the experimental results for TEM without and with energy filtering in Figs. S4 and S5 vs. Fig. 5, respectively.

Since $\ell_{\mathrm{mfp}}$ varies with both material and imaging conditions, it is challenging to calculate $\ell_{\mathrm{mfp}}$ theoretically for any specimen. Hence, we should determine $\ell_{\mathrm{mfp}}$ from a BF-TEM image of a calibrated specimen with the same material with the same imaging conditions. Since we knew the thickness of a particular region in the specimen (e.g., the substrate thickness), we were able to use Eq. (7) to determine the value of $\ell_{\mathrm{mfp}}$. To minimize spurious thickness changes due to spatial variations in beam intensity and detector response, a reference TEM image without any specimen, $I_0(x,y)$, was captured at the same imaging conditions as $I_\mathrm{t}(x,y)$.

## Depth estimation for multi-layer samples

For multi-layered specimens, the thickness information cannot be readily resolved for each layer. However, the additive CTF model in Eq. (8) shows that we can infer the depth information of each layer if both layers are not too thick.

$$|I_{\mathrm{det}}(\mathbf{k})| \approx A_1 \, \mathrm{env}_1(k) \, |\mathrm{CTF}_1(\mathbf{k})| + A_2 \, \mathrm{env}_2(k) \, |\mathrm{CTF}_2(\mathbf{k})| + \mathrm{noise}(k), \tag{8}$$

where $A_n$, $\mathrm{CTF}_n(\mathbf{k})$, and $\mathrm{env}_n(k)$ are amplitude, CTF and envelope functions for the corresponding top (n=1) and bottom (n=2) layers. Here, the CTF and envelope terms are related to the terms in Eqs. (2-4).

Since this is an additive model, the sum of additive noise terms from each layer can be combined into one term as defined in Eq. (5). We compare the radial profile of Thon rings from a dual-layer specimen and the additive model (Eq. (8)) in Fig. 4a and this additive model is used to generate the dual-layer structure in Fig. 4b and 4c. The amplitudes $A_n$ from the additive model can be used in estimating the relative thickness between the layers, if we have a large enough patch (400x400 pixels) and a high enough dose (~20k e Å$^{-2}$). The relative thickness estimated from amplitudes is used to generate the dual-layer structure in Fig. 4c.

The two-layer cases shown in Fig. 4b and 4c demonstrated the practical efficacy of the additive-CTF model: essentially, the interference of the wavefunction between the different layers can be ignored with regard to pop-out. Hence, the method can be extended to multi-layered specimens. However, determining the practical upper limit on the number of layers for realistic samples is not straightforward. In general, this limit depends on the number of distinct Thon rings from each layer. This distinguishability criterion, in turn, depends on the sample's total optical thickness, focal range, and signal.

### Running window averages of sample thickness and depth

In principle, the thickness value at every pixel in the image ($xy$ plane) can be used to pop out material symmetrically along the $z$ axis on either side of the centre of the mass value of that pixel (defocus map). Recall that we can only compute the average defocus for each image patch; hence this creates a resolution gap between our estimates for thickness versus depth. In practice, this gap is smaller because we compute a more noise-robust average thickness over a relatively small multi-pixel window. This is done for a more noise-robust estimate of the sample thickness from Eq. (7). Nevertheless, the side length of the patches used for CTF fitting is still larger than those of the windows used to estimate the average sample thickness. For example, in Fig. 5a, we used a CTF-fitting patch size of 30 nm (300 pixels), while average thickness windows of 7.5 nm (75 pixels). Hence, the former sets the conservative $xy$ resolution of our pop-out reconstructions.

For similar noise-robustness in our estimates of sample depth, we also computed the average defocus map $\Delta f'_{(i,j)}$ with overlapping windows with a stride length less than the patch size. In Fig. 5a, the stride length of 7.5 nm (75 pixels) for the 30 nm patch size provides three overlapping patches between any two non-overlapping patches. Choosing the stride size similar to the thickness window size resolves the resolution gap issue. Nevertheless, we can further determine the resultant running-window average defocus values for every $(x, y)$ pixel in the image from Eq. (9).

$$\Delta f_{(x,y)} = \frac{\sum_i \sum_j \Delta f'_{(i,j)} V_{(i,j,x,y)}}{\sum_i \sum_j V_{(i,j,x,y)}}, \qquad (9)$$

where $\Delta f'_{(i,j)}$ is the average defocus calculated from a particular window indexed by $(i,j)$. $V_{(i,j,x,y)}$, the visitation weights, is an array of ones and zeros: array element $V_{(i,j,x,y)}$ takes on the value of one only if pixel $(x, y)$ is visited by a particular window indexed by $(i,j)$.

Although the defocus map $\Delta f_{(x,y)}$ can be calculated for each pixel, the pixels within a stride length have the same defocus value because they all are calculated from the same sets of windows. The red square in Fig. 6a shows one such group of pixels that are within a stride length (10 nm) and is covered by three overlapping patches (cyan, blue, and magenta squares with dashed lines, each with 30 nm side length). If we defined the position of this red square as $(x, y)$, then the sum of the visitation weights $\sum_i \sum_j V_{(i,j,x,y)}$ for all such stride-length regions in the entire micrograph is shown in Fig. 6b.

Apart from the shot noise, the crosstalk between the phase contrast and amplitude contrast affects the thickness and defocus determination. The phase contrast produces light-dark fringes, and these fringes are more prominent near the sharp edges of the specimen. Thus, the number of electrons in the image pixels near these edges does not correspond to the material thickness. Similarly, the defocus map would be affected near such edges in the specimen as the defocus values within the window change abruptly. As a workaround, the fit-error for the defocus parameter in the CTF fitting can be used to determine which defocus values are erroneous (fit-error above a certain threshold) and are discarded from the visitation weights $V_{(i,j,k,l)}$. In Fig. 6d, the defocus and thickness values of the red region are calculated from votes from the three patches (cyan, blue, and magenta). From Fig. 1b, this red region straddles an abrupt change in the 3D structure. However, only the blue and magenta patches include this abrupt structural change, which leads to them having substantially larger defocus fit errors (see insets in Fig. 6a) compared to the cyan patch (which does not see this abrupt structural change). Since the lateral resolution is set by the patch size of 30nm side length, the defocus and thickness in the red region are only informed by its overlapping patches whose defocus fit errors (used to determine axial resolution) are below 30nm. Hence, the defocus and thickness values of the red region are averaged over the values of the qualified patches, as shown in Fig. 6e. The updated visitation weights in Fig. 6c indicate that only one patch was qualified for the red region. Fig. 6d shows the unweighted mean defocus and thickness values, while Fig. 6e shows the visitation-weighted mean defocus and thickness. The zero qualified visitation regions in Fig. 6c cause gaps in the final defocus map $\Delta f_{(x,y)}$, which are then filled by the nearest neighbour interpolation scheme. First, the regions with non-zero-sum of the qualified visitation weights are filled by the weighted mean defocus and thickness values. Then the remaining regions with zero-sum of the qualified visitation weights would get the weighted mean defocus and thickness values of the nearest non-zero neighbour stride-length region.

With defocus map $\Delta f_{(x,y)}$ and thickness map $T_{(x,y)}$ at reconstruction resolution, a 3D model is obtained using Eq. (10). To obtain an isometric reconstruction, either the defocus and thickness values should be scaled to match their $xy$ reconstruction voxel size, or the reconstruction voxel size should be set to match the units of their values ($z$ axis). Otherwise, the reconstruction would be anisometric.

$$P_{(x,y,z)} = \begin{cases} 1, & \Delta f_{(x,y)} - \frac{T_{(x,y)}}{2} \leq z \leq \Delta f_{(x,y)} + \frac{T_{(x,y)}}{2}, \\ 0, & \text{else} \end{cases} \qquad (10)$$

For thicker specimens, the specimen-depth-based change in effective magnification should be accounted for in the reconstruction, especially when there is a significant change (>10%) between the top and bottom surface of the specimen (Eq. (44)).

## Implementation of pop-out metrology

The linear regression for depth estimates in Eq. (2) is performed using the Levenberg-Marquardt algorithm implemented in the SciPy package [56]. The total computation time required for the pop-out 3D metrology reconstruction is based on the search space of the CTF fitting, TEM image size, window size, and step size (for overlapping windows). The knowledge of predetermined ranges of defocus and astigmatism values accelerates our parameter regressions by searching in a much smaller parameter space. When ranges of defocus and other parameters are unknown, unbounded fitting can be performed for a single or few characteristic patches from the image to determine them so that the parameter search space for the entire TEM image can be narrowed down. Since we normalize the spectra, the bounds for amplitude are set to a low non-zero value and one. We use 5 to 20% bounds for the amplitude contrast ratio, commonly used in the cryo-EM community for CTF-fitting[41,57]. The empirical method for finding the bounds for the rest of the fitting parameters is described below.

1. First, we do a coarse grid search on these few characteristic patches to find the defocus range.
2. Then, we use the appropriate defocus values for these patches to initialize unbounded fitting to fine-tune the bounds for envelope and noise parameters.
3. These bounds are set for the 1D fitting of all patches. The respective fit variances of the parameters from the 1D fitting are used to set the bounds in the 2D fitting.

Should the determined astigmatism be negligible in these initial fits, we can avoid 2D fitting to speed up parameter regressions for other patches (see Supplementary Software 1).

## Derivation of semi-empirical multislice scattering

In the multislice scheme (shown in Fig. 7a), a slab of homogeneous scattering material of thickness $T$ is partitioned into N thin slabs along the optical z-axis, each of thickness $\Delta z = T/(N-1)$. The thinness of each slab allows the scattering potential of its constituent atoms to be projected to a single infinitesimally thin two-dimensional (2D) slice. Such a projection approximation effectively turns N slabs into N 2D slices.

The multislice scheme alternately applies two operations: (1) the scattering potential of each slice modifies the electron wavefunction that is incident upon it; (2) then a free-space propagator then propagates this modified wavefunction to the next slice, which in turn becomes the incident wavefunction for this next slice. This alternating operation takes the incident electron wavefunction from the first scattering slab through the final occupied slab. The exiting wavefunction from the final occupied slab is then propagated to the imaging plane, which includes the optical aberrations of the microscope's image-forming lenses.

Below, we will recast the scattering from multiple slices in the previous paragraph into that of a single effective slice. We start by considering the scattering contributions from each of these $N$ slices. In the weak phase approximation, the exit waves of electron plane waves of wavelength $\lambda$ after the first few of these slices (indexed $n = 1, 2, 3, ...$) are respectively

$$\psi_1(r) = \exp[i\, s_1(r)] \approx 1 + i\, s_1(r), \tag{11}$$
$$\psi_2(r) \approx [\psi_1(r) \otimes p_1(r)]\,(1 + i\, s_2(r)), \tag{ }$$
$$\psi_n(r) \approx [\psi_{n-1}(r) \otimes p_{n-1}(r)]\,(1 + i\, s_n(r)), \tag{12}$$

where $p_n(r)$ is the two-dimensional (2D) kernel function that propagates the wavefront from slice $n$ to slice $n+1$, and $\otimes$ is the 2D convolution operator; the scattering potential distribution of the $n$th slice is defined by

$$s_n(r) \equiv v_n(r) - i\, \mu_n(r) \approx v_n(r)(1 + i\,\epsilon), \tag{13}$$

where $v_n$ and $\mu_n$ are the real and imaginary parts of the $n$th-slice's z-projected scattering potential. Here, we make the approximation that the real and imaginary parts of the scattering potential are related via a multiplicative constant $\epsilon$, which correlates to the inelastic scattering of the specimen.
For sufficiently thin slices, terms of order $|s_n|^2$ can be ignored, the exit wave after $N$ slices can be generalized as

$$\psi_N(r) \approx 1 + \sum_{n=1}^{N} i\, s_n(r) \otimes p_{N-n}(r), \tag{14}$$

where $p_{N-n}(r)$ is the propagator to advance the exit wave through $N-n$ slices. The result in Eq. (14) essentially ignores multiple scattering and only accounts for the fact that the exit wave from farther slices must be propagated over longer distances to match up with the exit wave at the final *N*-th slice.

We denote $\psi_{\text{det}}(r)$ as the wavefunction that is incident on the image-forming detector, which includes aberrations due to post-sample optical elements (i.e., objective lens, etc.). The 2D Fourier transform of this wavefunction is

$$\Psi_{\text{det}}(\mathbf{k}) \approx \delta(\mathbf{k}) + \sum_{n=1}^{N} i \exp[-i\chi(k)] S_n(\mathbf{k}) P_{N-n}(\mathbf{k}), \tag{15}$$

with $S_n(\mathbf{k})$ as the Fourier transform of $s_n(x)$; the Fourier transform of the propagator $p_{N-n}(x)$ is

$$P_{N-n}(\mathbf{k}) \equiv \exp(i(N-n)\theta) \equiv \exp[i(N-n)\pi\lambda k^2 \Delta z], \text{ where } \theta \equiv \pi\lambda k^2 \Delta z \tag{16}$$

and the aberration function in the post-sample image-forming lenses is

$$\chi(k) \equiv \frac{2\pi}{\lambda}\left[\frac{C_s}{4}\lambda^4 k^4 - \frac{1}{2}(\Delta f)\lambda^2 k^2\right], \quad \text{with } k \equiv |\mathbf{k}|, \tag{17}$$

With $\Delta f$ as the relative defocus of the final *N*-th slice from the plane of focus, and $C_s$ as the spherical aberration parameter of the microscope's image-forming lenses.

Now, if we defined $\phi_n(\mathbf{k}) = S_n(\mathbf{k})\exp(-in\theta)$, and $\tilde{\chi}(\mathbf{k}) = \chi(k) - N\theta$, then we can rewrite Eq. (15) as

$$\Psi_{\text{det}}(\mathbf{k}) \approx \delta(\mathbf{k}) + i\exp(-i\tilde{\chi}(k)) \sum_{n=1}^{N} \phi_n(\mathbf{k}). \tag{18}$$

Hence, the probability of detecting electrons on the detector is (when dropping terms of order $\phi^2$ or higher because the scattering from each thin slice is small), which is measured as intensities on the detector

$$I_{\text{det}}(r) \equiv |\psi_{\text{det}}(r)|^2 \approx 1 + i h(r) \otimes \sum_{n=1}^{N} \phi_n(r) - i h^*(r) \otimes \sum_{n=1}^{N} \phi_n^*(r). \tag{19}$$

where $h(r)$ is the point spread function in real space, i.e., the Fourier transform of $\exp(-i\tilde{\chi}(k))$.

Fourier transforming this intensity gives

$$I_{\text{det}}(\mathbf{k}) = \delta(\mathbf{k}) + i \sum_{n=1}^{N} \left[\phi_n(\mathbf{k})\exp(-i\tilde{\chi}(k)) - \phi_n^*(-\mathbf{k})\exp(i\tilde{\chi}(k))\right], \tag{20}$$

where $\delta(\mathbf{k})$ is the Dirac delta function. Using the approximation from Eq (13) into $\phi_n(\mathbf{k})$ and $\phi_n^*(-\mathbf{k})$ in Eq. (20), we obtain,

$$I_{\text{det}}(\mathbf{k}) = \delta(\mathbf{k}) + 2\sqrt{1+\epsilon^2} \sum_{n=1}^{N} v_n(\mathbf{k}) \sin(\tilde{\chi} + n\theta - \alpha), \quad \text{where } \alpha = \arctan(\epsilon). \tag{21}$$

As an instructive curiosity, we can make the *rather unphysical assumption* that all slices are identical (i.e., $v_n(\mathbf{k}) = v(\mathbf{k})$), although $v(\mathbf{k})$ itself is random. In this case, we can pull out the $v(\mathbf{k})$ term from the sum in Eq. (21), which can be rewritten as

$$I_{\text{det}}(\mathbf{k}) = \delta(\mathbf{k}) + 2\sqrt{1+\epsilon^2}\, v(\mathbf{k}) \sum_{n=1}^{N} \sin\left(\tilde{\chi} - \alpha + \left(n - \frac{N}{2}\right)\theta\right), \tag{22}$$

where $\bar{\chi} = \chi - N\theta/2$ is the average aberration function as measured from the middle slice (i.e., $n = N/2$). The summation in Eq. (22) can be approximated as an integral (assuming sufficiently thin slices, $\Delta z \to 0$, see Fig. 7b) to give

$$I_{\text{det}}(\boldsymbol{k}) \approx \delta(\boldsymbol{k}) + 2\sqrt{1+\epsilon^2}\, \nu(\boldsymbol{k}) \int_{-T/2}^{T/2} \sin(\bar{\chi} - \alpha + z\pi\lambda k^2)\, dz. \tag{23}$$

Resolving the integral in Eq. (23) gives

$$I_{\text{det}}(\boldsymbol{k}) \approx \delta(\boldsymbol{k}) + 2\sqrt{1+\epsilon^2}\,(T\,\nu(\boldsymbol{k}))\sin(\bar{\chi} - \alpha)\frac{\sin(\xi/2)}{\xi/2}, \qquad \text{where } \xi = \pi\lambda k^2 T. \tag{24}$$

The resultant power spectrum from Eq. (24) then becomes

$$|I_{\text{det}}(\boldsymbol{k})|^2 = \delta(\boldsymbol{k}) + 2\,(1+\epsilon^2)\,|T\nu(\boldsymbol{k})|^2 \bigl(1 - \cos(2(\bar{\chi} - \alpha))\bigr)\left(\frac{\sin(\xi/2)}{\xi/2}\right)^2. \tag{25}$$

Critically, the $\cos(2(\bar{\chi} - \alpha))$ term in Eq. (25) clearly shows how the effective defocus of the entire sample is now centred at the centre of the scattering mass of the sample (i.e., $z = T/2$) as shown in Fig. 7. This conclusion was first observed by Bonhomme *et al.*[44]

This unphysical (identical, random slice) assumption leads to the nodes of the squared sinc function in Eq. (25) to occur at

$$\frac{\xi}{2} = j\pi : j \in \mathbb{Z}^+, \qquad \text{or} \qquad \frac{\lambda k^2 T}{2} = 1,2,3, \dots, \tag{26}$$

which is the result first obtained by Bonhomme *et al.*[44] These node positions, however, have been later shown by Tichelaar *et al.* to be incorrect using tomography.[58]

If instead we assume that the more realistic scenario where the scattering potential of different slices $\nu_n(\boldsymbol{k})$ are random and different, the power spectrum in Eq. (21) now becomes

$$|I_{\text{det}}(\boldsymbol{k})|^2 = 4\,(1+\epsilon^2)\,(\,\mathrm{B}(\boldsymbol{k}) + \Gamma(\boldsymbol{k})\,), \tag{27}$$

$$\mathrm{B}(\boldsymbol{k}) = \sum_{n=1}^{N} \beta_n(\boldsymbol{k}),$$
$$\mathrm{B}(\boldsymbol{k}) = \sum_{n=1}^{N} |\nu_n(\boldsymbol{k})|^2 \sin^2(\tilde{\chi} + n\theta - \alpha), \tag{28}$$

$$\Gamma(\boldsymbol{k}) = \sum_{n=1}^{N}\sum_{m>n}^{N} \gamma_{mn}(\boldsymbol{k}),$$
$$\Gamma(\boldsymbol{k}) = \sum_{n=1}^{N}\sum_{m>n}^{N} 2\,\mathrm{Re}[\nu_n(\boldsymbol{k})\nu^*_m(-\boldsymbol{k})]\,\sin(\tilde{\chi} + n\theta - \alpha)\sin(\tilde{\chi} + m\theta - \alpha),$$
$$\Gamma(\boldsymbol{k}) = \sum_{n=1}^{N}\sum_{m>n}^{N} 2\,\mathrm{Re}[\nu_n(\boldsymbol{k})\nu_m(\boldsymbol{k})]\,\sin(\tilde{\chi} + n\theta - \alpha)\sin(\tilde{\chi} + m\theta - \alpha). \tag{29}$$

The simplification in the last step comes about because the scattering potential $\nu(\boldsymbol{r})$ in Eq. (13) is real-valued, hence its Fourier transform is centro-symmetric: $\nu_m(\boldsymbol{k}) = \nu^*_m(-\boldsymbol{k})$.

To make progress, we approximated $\nu_n(\boldsymbol{k})$ by creating thin slices of random $SiN_x$. Fig. S1 shows that the cross multiplication term $\Gamma(\boldsymbol{k})$ is much smaller compared to the term $\mathrm{B}(\boldsymbol{k})$. Fig. S1 also shows the one-dimensional angular average of the power spectrum $\langle |I_{\text{det}}(\boldsymbol{k})|^2 \rangle_{|\boldsymbol{k}|=k}$. Using such random slices in Eqs. (27-29), we see that the sinc-like nodes of the angularly averaged power spectrum occur when

$$\lambda k^2 T = 1, 2, \ldots \in \mathbb{Z}. \tag{30}$$

The node positions in Eq. (30) are consistent with those shown in Tichelaar *et al.*, which were experimentally validated by the authors using tomography[58]. These same node positions were also proposed by McMullan *et al.* but with less rigor than the mathematical exposition presented in this section.[59]

Importantly, even for the random $v_n(\mathbf{k})$ case, the effective defocus of the entire sample in $\bar{\chi}$ is still centred at the centre of the scattering mass of the sample (i.e., $z = T/2$). This has been verified in the multislice simulations in Fig. S1.

Supplementary Note 1 shows that $\Gamma(\mathbf{k})$ term is numerically small, and CTF undulations follow $\sin^2(\bar{\chi} - \alpha)$ with a sinc-like envelope caused by the specimen thickness. Apart from the specimen thickness, there are a plethora of effects such as spatial and temporal incoherence, specimen motion, charging effects, beam-induced movement, and stage-drift contribute to the envelope function. Hence, a Gaussian envelope can be used as a cumulative envelope function in the model.[55] The validation for choosing a Gaussian over a sinc function for the envelope is provided in Supplementary Note 2.

Since $\Gamma(\mathbf{k})$ term is small and does not modify the undulations and the node positions, we can rewrite Eq. (27) as

$$|I_{\text{det}}(\mathbf{k})| \approx A\,\text{env}(k)\,|\sin(\bar{\chi} - \alpha)| + \text{noise}(k), \tag{31}$$

where $A$ is a multiplicative constant proportional to $\sqrt{1 + \epsilon^2}$ and combined multiplicative and additive terms dependent on $k$ are modelled with $\text{env}(k)$ and $\text{noise}(k)$ (Eqs. (4-5)). The depth information is encoded in the term $\sin(\bar{\chi} - \alpha)$, which corresponds to the CTF function.

$$\sin(\bar{\chi} - \alpha) = \sin(\bar{\chi})\cos(\alpha) - \sin(\alpha)\cos(\bar{\chi}). \tag{32}$$

Since $\alpha = \arctan(\epsilon)$, $\cos(\alpha) = \frac{1}{\sqrt{1+\epsilon^2}}$ and $\sin(\alpha) = \frac{\epsilon}{\sqrt{1+\epsilon^2}}$. To be consistent with the literature, the coefficient of cosine term is written as the amplitude contrast ratio $Q$, i.e., $Q = \frac{\epsilon}{\sqrt{1+\epsilon^2}}$ and $\sqrt{1 - Q^2} = \frac{1}{\sqrt{1+\epsilon^2}}$, then we can rewrite Eq. (32) as

$$\sin(\bar{\chi} - \alpha) = \sqrt{1 - Q^2}\sin(\bar{\chi}) - Q\cos(\bar{\chi}). \tag{33}$$

Substituting this in Eq. (31) gives our CTF model shown in Eqs. (2-3)

$$|I_{\text{det}}(\mathbf{k})| \approx A\,\text{env}(k)\,|\text{CTF}(\mathbf{k})| + \text{noise}(k),$$

$$\text{CTF}(\mathbf{k}) = w_1 \sin(\bar{\chi}) - w_2 \cos(\bar{\chi}).$$

**Resolution limit for a thick sample**

The patch size for the CTF fitting (see Eq. 3) defines the $xy$ resolution of the pop-out 3D metrology. Many parameters determine the patch size, including TEM image resolution. Increasing the TEM image resolution increases the reconstruction resolution. However, for thicker samples, sinc-like nodes (Fig. S1), as mentioned above, limit our reconstruction resolution. Due to incoherence and noise, it is hard to obtain clear undulations beyond the first node (which occurs at $\lambda k^2 T = 1$). Since the resolution and the thickness are inversely proportional in determining the node position, the lower bound of our transverse resolution limit

$k_{\lim}$ for our reconstruction is set by the thickest part of the sample ($T_{\max}$), regardless of the defocus parameter $\Delta f$:

$$\lambda k_{\lim}^2 T_{\max} = 1, \tag{34}$$

$$k_{\lim} = \sqrt{1/\lambda T_{\max}}, \tag{35}$$

where $T_{\max}$ is the thickness of the thickest region of the specimen along the optical axis. The TEM's magnification should be chosen correspondingly to achieve an image's resolution greater than $k_{\lim}$. This CTF fitting should be able to determine the defocus from both the nearest and the farthest points of the specimen. The point nearest to the focal plane, whose defocus parameter we denote as $\Delta f$, will have the fewest number of CTF rings in its image's Fourier transform. In contrast, the farthest point, which we denote here to have depth $\Delta f + \Delta f_r$ from the focal plane, would have the most number of CTF rings. The aberration parameter $\tilde{\chi}(k_{\lim}, \Delta f)$ (Eq. 17) of the nearest point at $k_{\lim}$ will be:

$$\Delta f \geq T_{\max} \left( \eta + \frac{C_s \lambda}{2 T_{\max}^2} \right), \tag{36}$$

where $\eta = \tilde{\chi}(k_{\lim}, \Delta f)/\pi$ is the number of CTF rings up to $k_{\lim}$. To paraphrase, this last equation gives us the minimum defocus $\Delta f$ needed to guarantee at least $\eta$ CTF rings up to $k_{\lim}$.

However, these CTF rings of a patch have to be finely sampled enough to determine the patch's average depth. This is equivalent to requiring that we satisfy a separate sampling criterion for the farthest point, which comprises $\eta_r$ CTF rings at $k_{\lim}$ (i.e., $\frac{\tilde{\chi}(k_{\lim}, \Delta f + \Delta f_r)}{\pi} = \eta_r$). In other words,

$$-\eta_r = \frac{C_s \lambda^3 k_{\lim}^4}{2} - \lambda k_{\lim}^2 (\Delta f + \Delta f_r). \tag{37}$$

The sampling criterion is that we have at least $\sigma_s$ frequency samples (spaced apart by $w^{-1}$ for patch sizes of side length $w_{\min}$ pixels) spanning between the $\eta_r^{\text{th}}$ and $(\eta_r - 1)^{\text{th}}$ CTF ring. Because the latter ring occurs at a spatial frequency

$$k_j = \sqrt{\frac{\lambda(\Delta f + \Delta f_r) - \sqrt{(\lambda(\Delta f + \Delta f_r))^2 - 2\lambda^3 C_s(\eta_r - 1)}}{\lambda^3 C_s}}, \tag{38}$$

this sampling criterion translates into at least having patches whose side lengths are

$$w_{\min} \geq \frac{\sigma_s}{\frac{1}{\sqrt{\lambda T_{\max}}} - k_j}. \tag{39}$$

**Extension to multiple layers.**

Eq. (27) can be trivially extended to apply to multiple layers. Here we consider the two layer case and leave the generalization to more than two layers to the motivated reader. The power spectrum of a two-layered sample can be written as

$$|I_{\text{det}}(\boldsymbol{k})|^2 = 4(1 + \epsilon^2)(B_1(\boldsymbol{k}) + B_2(\boldsymbol{k}) + \Gamma(\boldsymbol{k})) \propto B_1(\boldsymbol{k}) + B_2(\boldsymbol{k}) + \text{noise}, \tag{40}$$

where $B_1(\boldsymbol{k})$ and $B_2(\boldsymbol{k})$ are the partial sums in Eq. (28) of the material slices within layer 1 and layer 2, respectively. The $\Gamma(\boldsymbol{k})$ term in Eq (40) is identical to that Eq. (29) except it runs over pairs of slices in both layers, as well as pairs of slices within each layer. Nevertheless, as Fig. S1 shows, this $\Gamma(\boldsymbol{k})$ term should vanish for amorphous materials where the slices are random and uncorrelated. Overall, the respective depths of layers 1 and 2 can be separately inferred from the sum of their $B_1(\boldsymbol{k})$ and $B_2(\boldsymbol{k})$ terms, which separately describe the CTFs of layers 1 and 2 respectively.

**Change in effective magnification at different specimen depths.**

When the specimen is thicker or has multiple layers far apart in the axial direction, the features at different depths would be encoded at different effective magnifications in the micrograph. Here, we analyze the influence of this change in effective magnification concerning the specimen depth in pop-out reconstructions.

Magnification $M$ is related by the ratio of distances between the image plane and the focal plane $V$ to the distance between the specimen to the focal plane $U$:

$$M = \frac{-V}{U}. \tag{41}$$

The negative sign indicates that the image plane and the specimen are on opposite sides of the focal plane. Differentiating the Eq. (41) concerning $U$ gives

$$\frac{\partial M}{\partial U} = \frac{V}{U^2}. \tag{42}$$

By substituting $U$ from Eq. (41), Eq. (42) can be written as

$$\frac{\partial M}{\partial U} = \frac{M^2}{V}. \tag{43}$$

The relative change in magnification for a change in the specimen depth is given by

$$\frac{\partial M}{M} = \frac{M}{V} \partial U. \tag{44}$$

In the images in Fig.5, the specimens' thickness is 200nm, and the magnification is 60,000. The distance between the image plane and the focal plane $V$ would be several hundred millimeters to a few meters (say, 0.5m). Substituting these values in Eq. (44) gives us a 2.4% change in the magnification between the specimens' top and bottom surfaces ($\partial U$ = 200 nm). The change in voxel size from the top surface to the bottom is less than 1nm. Though such a small difference can be ignored in the reconstruction, when there is a significant change in magnification/voxel size, the reconstructed voxels need to be corrected accordingly.

## Data availability

Simulated datasets (Fig.1 and Fig. 2) for the purpose of demonstration of the code are available in Supplementary Software 1 and in Zenodo open data repository[60]. Other data underlying the results presented in this paper are not publicly available at this time but may be obtained from the authors upon reasonable request.

## Code availability

*Python* implementation of the algorithm with simulated datasets for reconstruction in (Fig.1 and Fig. 2) is available in Supplementary Software 1 and in Zenodo open data repository[60].

## References

1. Koumoulos, E. P. *et al.* Metrology and nano-mechanical tests for nano-manufacturing and nano-bio interface: Challenges & future perspectives. *Mater. Des.* **137**, 446–462 (2018).
2. Witold, L., Rasit, T., Ana, P. & Agnieszka, D. Eighth Nanoforum Report. (2006).
3. Miele, E., Raj, S., Baraissov, Z., Král, P. & Mirsaidov, U. Dynamics of Templated Assembly of Nanoparticle Filaments within Nanochannels. *Adv. Mater.* **29**, (2017).
4. Holler, M. *et al.* Three-dimensional imaging of integrated circuits with macro- to nanoscale zoom. *Nature Electronics* **2**, 464–470 (2019).
5. Witte, K. *et al.* From 2D STXM to 3D Imaging: Soft X-ray Laminography of Thin Specimens. *Nano Lett.* **20**, 1305–1314 (2020).
6. Sala, S. *et al.* Ptychographic X-ray computed tomography at a high-brilliance X-ray source. *Optics Express* vol. 27 533 Preprint at https://doi.org/10.1364/oe.27.000533 (2019).
7. Prakash, A. *et al.* Nanoscale cuticle density variations correlate with pigmentation and color in butterfly wing scales. *arXiv [physics.bio-ph]* (2023).
8. Conte, T. M. & Gargini, P. A. On the foundation of the new computing industry beyond 2020. *Preliminary IEEE RC-ITRS Report* (2015).
9. Orji, N. G. *et al.* Metrology for the next generation of semiconductor devices. *Nat Electron* **1**, (2018).
10. Kessler, R. M., Ellis, J. R., Jr & Eden, M. Analysis of emission tomographic scan data: limitations imposed by resolution and background. *J. Comput. Assist. Tomogr.* **8**, 514–522 (1984).
11. Goy, A. *et al.* High-resolution limited-angle phase tomography of dense layered objects using deep neural networks. *Proc. Natl. Acad. Sci. U. S. A.* **116**, 19848–19856 (2019).
12. Loh, N.-T. D. & Elser, V. Reconstruction algorithm for single-particle diffraction imaging experiments. *Phys. Rev. E Stat. Nonlin. Soft Matter Phys.* **80**, 026705 (2009).
13. Zhu, J., Penczek, P. A., Schröder, R. & Frank, J. Three-Dimensional Reconstruction with Contrast Transfer Function Correction from Energy-Filtered Cryoelectron Micrographs: Procedure and Application to the 70SEscherichia coliRibosome. *J. Struct. Biol.* **118**, 197–219 (1997).
14. Arslan, I., Yates, T. J. V., Browning, N. D. & Midgley, P. A. Embedded Nanostructures Revealed in Three Dimensions. *Science* **309**, 2195–2198 (2005).
15. Midgley, P. A. & Dunin-Borkowski, R. E. Electron tomography and holography in materials science. *Nat. Mater.* **8**, 271–280 (2009).
16. Van Aert, S., Batenburg, K. J., Rossell, M. D., Erni, R. & Van Tendeloo, G. Three-dimensional atomic imaging of crystalline nanoparticles. *Nature* **470**, 374 (2011).
17. Zhu, C. *et al.* Towards three-dimensional structural determination of amorphous materials at atomic resolution. *Phys. Rev. B Condens. Matter* **88**, 100201 (2013).
18. Ercius, P., Alaidi, O., Rames, M. J. & Ren, G. Electron Tomography: A Three-Dimensional Analytic Tool for Hard and Soft Materials Research. *Adv. Mater.* **27**, 5638–5663 (2015).
19. Hata, S. *et al.* Electron tomography: An imaging method for materials deformation dynamics. *Curr. Opin. Solid State Mater. Sci.* **24**, 100850 (2020).
20. Kwon, O.-H. & Zewail, A. H. 4D electron tomography. *Science* **328**, 1668–1673 (2010).
21. Kline, R. J. Multiscale 3D X-ray imaging. *Nature Electronics* **2**, 435–436 (2019).
22. Gabor, D. A new microscopic principle. *Nature* **161**, 777 (1948).
23. Schnars, U. & Jüptner, W. Direct recording of holograms by a CCD target and numerical reconstruction. *Appl. Opt.* **33**, 179–181 (1994).
24. Kim, M. K., Yu, L. & Mann, C. J. Interference techniques in digital holography. *J. Opt. A: Pure Appl. Opt.* doi:10.1088/1464-4258/8/7/S33.

25. Cowley, J. M., Moodie, A. F. & IUCr. The scattering of electrons by atoms and crystals. I. A new theoretical approach. *Acta Crystallogr.* **10**, 609–619 (1957).
26. Pang, S. & Barbastathis, G. Unified treatment of exact and approximate scalar electromagnetic wave scattering. *Phys Rev E* **106**, 045301 (2022).
27. Javidi, B. *et al.* Roadmap on digital holography [Invited]. *Opt. Express* **29**, 35078–35118 (2021).
28. Wang, H., Tahir, W., Zhu, J. & Tian, L. Large-scale holographic particle 3D imaging with the beam propagation model. *Opt. Express* **29**, 17159–17172 (2021).
29. Fung, J. *et al.* Measuring translational, rotational, and vibrational dynamics in colloids with digital holographic microscopy. *Opt. Express* **19**, 8051–8065 (2011).
30. Molaei, M. & Sheng, J. Imaging bacterial 3D motion using digital in-line holographic microscopy and correlation-based de-noising algorithm. *Opt. Express* **22**, 32119–32137 (2014).
31. Leahy, B., Alexander, R., Martin, C., Barkley, S. & Manoharan, V. N. Large depth-of-field tracking of colloidal spheres in holographic microscopy by modeling the objective lens. *Opt. Express* **28**, 1061–1075 (2020).
32. Brady, D. J., Choi, K., Marks, D. L., Horisaki, R. & Lim, S. Compressive holography. *Opt. Express* **17**, 13040–13049 (2009).
33. Hahn, J., Lim, S., Choi, K., Horisaki, R. & Brady, D. J. Video-rate compressive holographic microscopic tomography. *Opt. Express* **19**, 7289–7298 (2011).
34. Devaney, A. J. Nonuniqueness in the inverse scattering problem. *J. Math. Phys.* **19**, 1526–1531 (1978).
35. Kahl, F. & Rose, H. Theoretical Concepts of Electron Holography. in *Advances in Imaging and Electron Physics* 197–257 (Elsevier, 1995).
36. McMullan, G., Faruqi, A. R., Clare, D. & Henderson, R. Comparison of optimal performance at 300keV of three direct electron detectors for use in low dose electron microscopy. *Ultramicroscopy* **147**, 156–163 (2014).
37. Levin, B. D. A. Direct detectors and their applications in electron microscopy for materials science. *J. Phys. Mater.* **4**, 042005 (2021).
38. Malis, T., Cheng, S. C. & Egerton, R. F. EELS log-ratio technique for specimen-thickness measurement in the TEM. *J. Electron Microsc. Tech.* **8**, 193–200 (1988).
39. Kirkland, E. J. *Advanced Computing in Electron Microscopy*. (Springer Science & Business Media, 2013).
40. Thon, F. Zur Defokussierungsabhängigkeit des Phasenkontrastes bei der elektronenmikroskopischen Abbildung. *Zeitschrift Naturforschung Teil A* **21**, 476–478 (1966).
41. Mindell, J. A. & Grigorieff, N. Accurate determination of local defocus and specimen tilt in electron microscopy. *J. Struct. Biol.* **142**, 334–347 (2003).
42. Rohou, A. & Grigorieff, N. CTFFIND4: Fast and accurate defocus estimation from electron micrographs. *J. Struct. Biol.* **192**, 216–221 (2015).
43. Zhang, K. Gctf: Real-time CTF determination and correction. *J. Struct. Biol.* **193**, 1–12 (2016).
44. Bonhomme, P. & Beorchia, A. The specimen thickness effect upon the electron microscope image contrast transfer of amorphous objects. *Journal of Physics D: Applied Physics* **16**, 705 (1983).
45. D. Balakrishnan, S. W. Chee, M. Bosman, and D. Loh. TEM-based single-shot coherent 3D imaging. in *Imaging and Applied Optics Congress 2022 (3D, AOA, COSI, ISA, pcAOP)* (Optica Publishing Group, 2022, 2022).
46. Coene, W., Janssen, G., Op de Beeck M & Van Dyck D. Phase retrieval through focus variation for ultra-resolution in field-emission transmission electron microscopy. *Phys. Rev. Lett.* **69**, 3743–3746 (1992).
47. Hovden, R., Xin, H. L. & Muller, D. A. Extended depth of field for high-resolution scanning transmission electron microscopy. *Microsc. Microanal.* **17**, 75–80 (2011).

48. E, H. *et al.* Probe integrated scattering cross sections in the analysis of atomic resolution HAADF STEM images. *Ultramicroscopy* **133**, 109–119 (2013).
49. Xin, H. L. & Muller, D. A. Aberration-corrected ADF-STEM depth sectioning and prospects for reliable 3D imaging in S/TEM. *J. Electron Microsc.* **58**, 157–165 (2009).
50. Jungjohann, K. L., Evans, J. E., Aguiar, J. A., Arslan, I. & Browning, N. D. Atomic-scale imaging and spectroscopy for in situ liquid scanning transmission electron microscopy. *Microsc. Microanal.* **18**, 621–627 (2012).
51. Kim, M. J., Wanunu, M., Bell, D. C. & Meller, A. Rapid Fabrication of Uniformly Sized Nanopores and Nanopore Arrays for Parallel DNA Analysis. *Adv. Mater.* **18**, 3149–3153 (2006).
52. Wagner, F. R. *et al.* Preparing samples from whole cells using focused-ion-beam milling for cryo-electron tomography. *Nat. Protoc.* **15**, 2041–2070 (2020).
53. Feist, A., Rubiano da Silva, N., Liang, W., Ropers, C. & Schäfer, S. Nanoscale diffractive probing of strain dynamics in ultrafast transmission electron microscopy. *Struct Dyn* **5**, 014302 (2018).
54. Schwartz, J. *et al.* Real-time 3D analysis during electron tomography using tomviz. *Nat. Commun.* **13**, 4458 (2022).
55. Ludtke, S. J. & Chiu, W. Image restoration in sets of noisy electron micrographs. in *Proceedings IEEE International Symposium on Biomedical Imaging* 745–748 (2002).
56. Jones, E., Oliphant, T., Peterson, P. & Others. SciPy: Open source scientific tools for Python. Preprint at http://www.scipy.org/ (2001-).
57. Humphreys, C. J. & Hirsch, P. B. Absorption parameters in electron diffraction theory. *The Philosophical Magazine: A Journal of Theoretical Experimental and Applied Physics* **18**, 115–122 (1968).
58. Tichelaar, W., Hagen, W. J. H., Gorelik, T. E., Xue, L. & Mahamid, J. TEM bright field imaging of thick specimens: nodes in Thon ring patterns. *Ultramicroscopy* vol. 216 113023 Preprint at https://doi.org/10.1016/j.ultramic.2020.113023 (2020).
59. McMullan, G., Vinothkumar, K. R. & Henderson, R. Thon rings from amorphous ice and implications of beam-induced Brownian motion in single particle electron cryo-microscopy. *Ultramicroscopy* **158**, 26–32 (2015).
60. Balakrishnan, D. *et al. Single-shot, coherent, pop-out 3D metrology*. (2023). doi:10.5281/zenodo.8303793.

## Acknowledgements

This work was supported by the Singapore National Research Foundation's Competitive Research Program funding (NRF-CRP16-2015-05), National University of Singapore Early Career Research Grant awarded to N.D.L., and the Ministry of Education (MOE) Academic Research Fund (AcRF) Tier 1 grant (A-0004462-00-00). The authors would like to thank Jian Shi for pointing out the higher-order aberrations causing deviations in the depth determination, the IT administrators at the National University of Singapore Centre for Bio-imaging Sciences for their computing support, and Paul T. Matsudaira for insightful discussions.

## Author contributions

DB and NDL conceived the idea for this work with the help of MB and UM. NDL and DB developed the theory. DB wrote the code for the simulation and developed/implemented the algorithm under NDL's advice. SWC, ZB, and UM provided the samples. DB, SWC, and ZB collected the experimental micrographs. DB and NDL wrote the manuscript with contributions from all authors.

## Competing Interests

The authors declare no competing interests.

## Supplemental information

See the supplementary information document for supporting content.

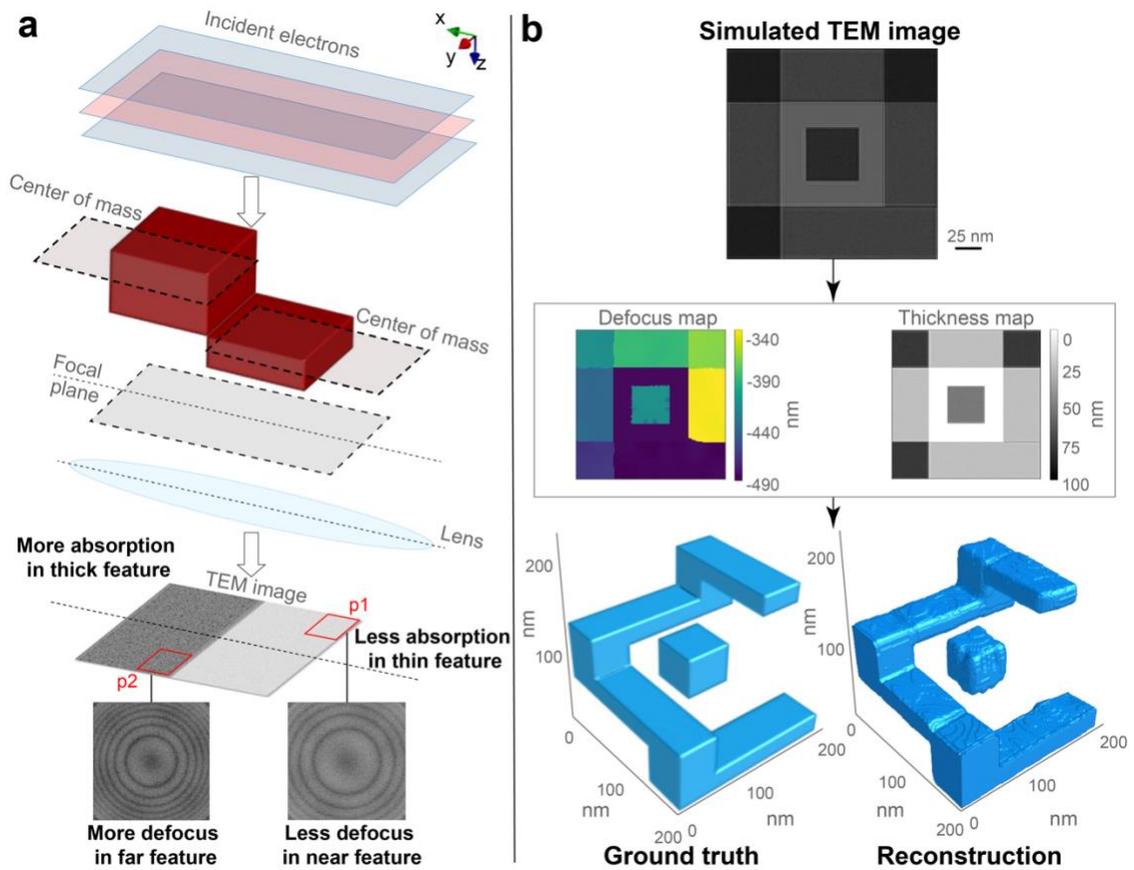

**Fig 1. Illustrating the principle of pop-out 3D metrology with transmission electron microscopy (TEM).** a) Schematic of bright-field TEM (BF-TEM). A thinner feature (right half) scatters fewer electrons and forms the brighter right half of the TEM image; it is also placed nearer to the focal plane. Hence its contrast transfer function (CTF) has fewer Thon rings than the thicker feature, as we can see from the Fourier transforms of the patches p1 and p2 (red squares). b) Applying the pop-out metrology technique to a 2048×2048 pixel simulated BF-TEM image of a 3D model (ground truth). The recovered average longitudinal centre of mass (defocus map) and the sample thickness map shown in the image were used to reconstruct the 3D volume.

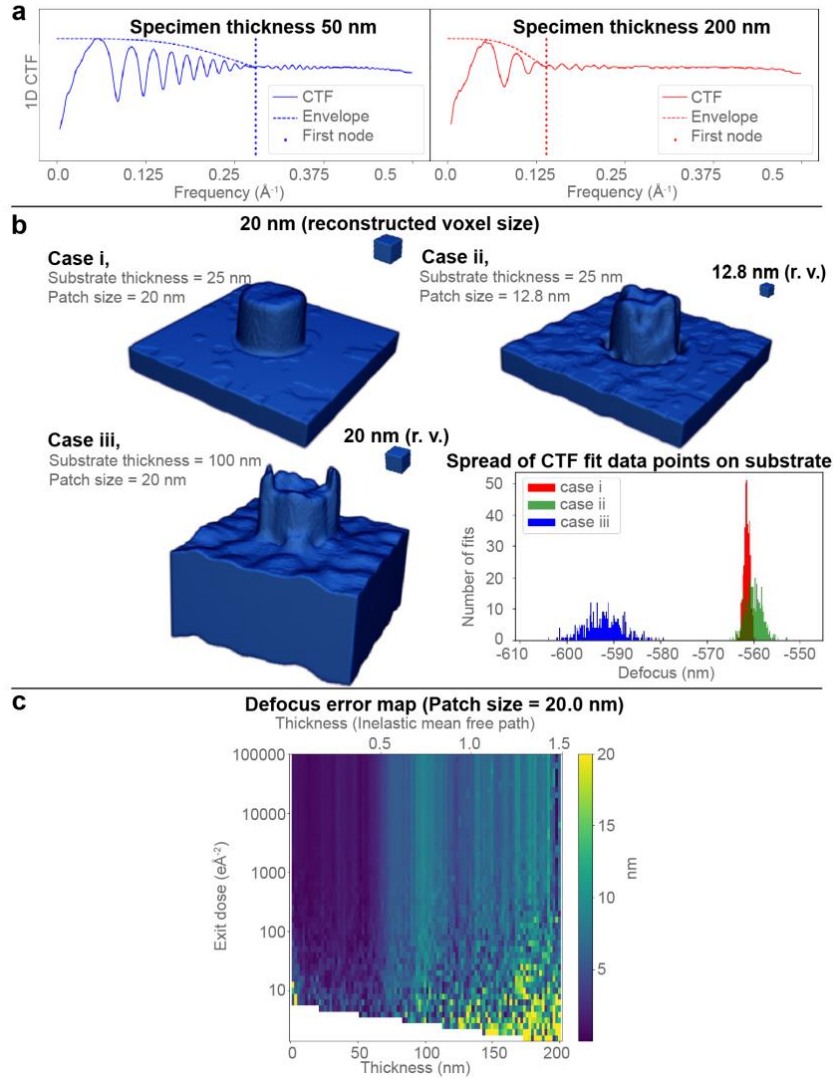

**Fig 2. Factors that limit in-plane (*xy*) and out-of-plane (*z*) resolutions.** a) Plots show the first zero-crossing of the envelope for thin and thick amorphous silicon nitride specimens. The thicker specimen has a steeper envelope which limits the resolution; hence it needs finer sampling of the spatial frequencies to fit the contrast transfer function (CTF) accurately. b) 3D reconstructions with the case i) optimal defocus-fit patch size (20 nm) for the given thickness (substrate 25 nm and pillar 50 nm), case ii) insufficient patch size (12.8 nm) for the same thickness, and case iii) 20 nm patch size for a thicker specimen (substrate 100 nm and pillar 50 nm). The histograms show the spread in defocus values in each case, i.e., the CTF-fitting precision. The same patch size, which was optimal in case i), is insufficient for a thicker specimen (case iii) as expected. c) The error between the fitted and actual defocus is plotted here as a function of exit dose and sample thickness (given patch size of 20 nm); Increasing the dose improves the defocus fit accuracy significantly at lower doses until 100 e Å$^{-2}$ exit dose. Increasing the patch size helps to sample the frequencies finer in the Fourier space; hence it improves the accuracy of depth fitting. As the patch size (*xy* resolution) is sufficiently large (20 nm), the accuracy value (*z* resolution) stays below the patch size.

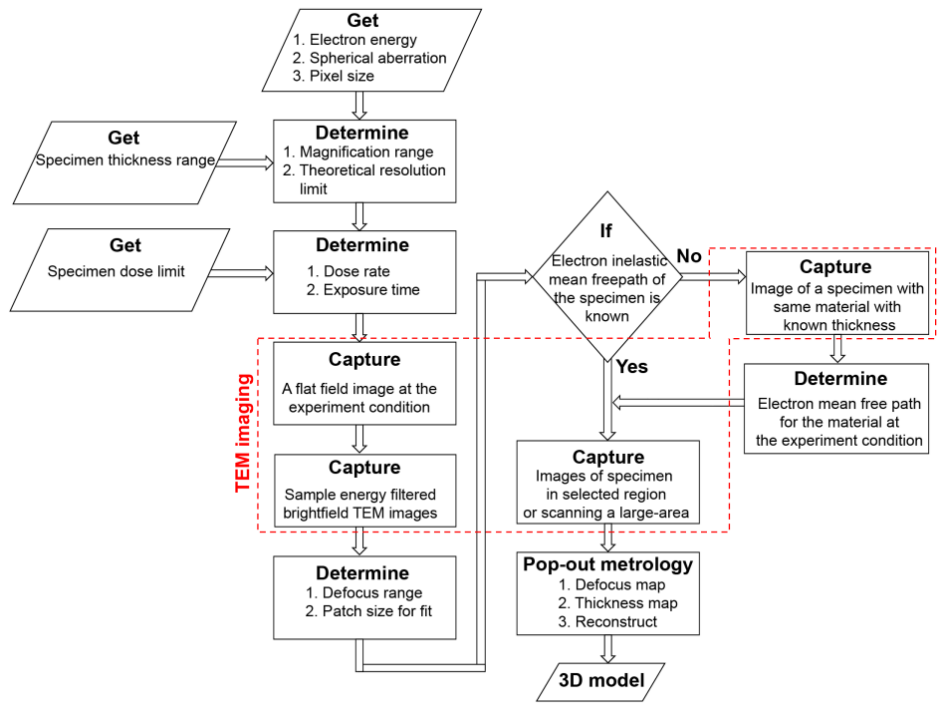

**Fig 3. Process flowchart for pop-out metrology.** A process flow chart that explains the step-by-step process and necessary inputs required at various steps for pop-out metrology. The dashed red box indicates the data acquisition process.

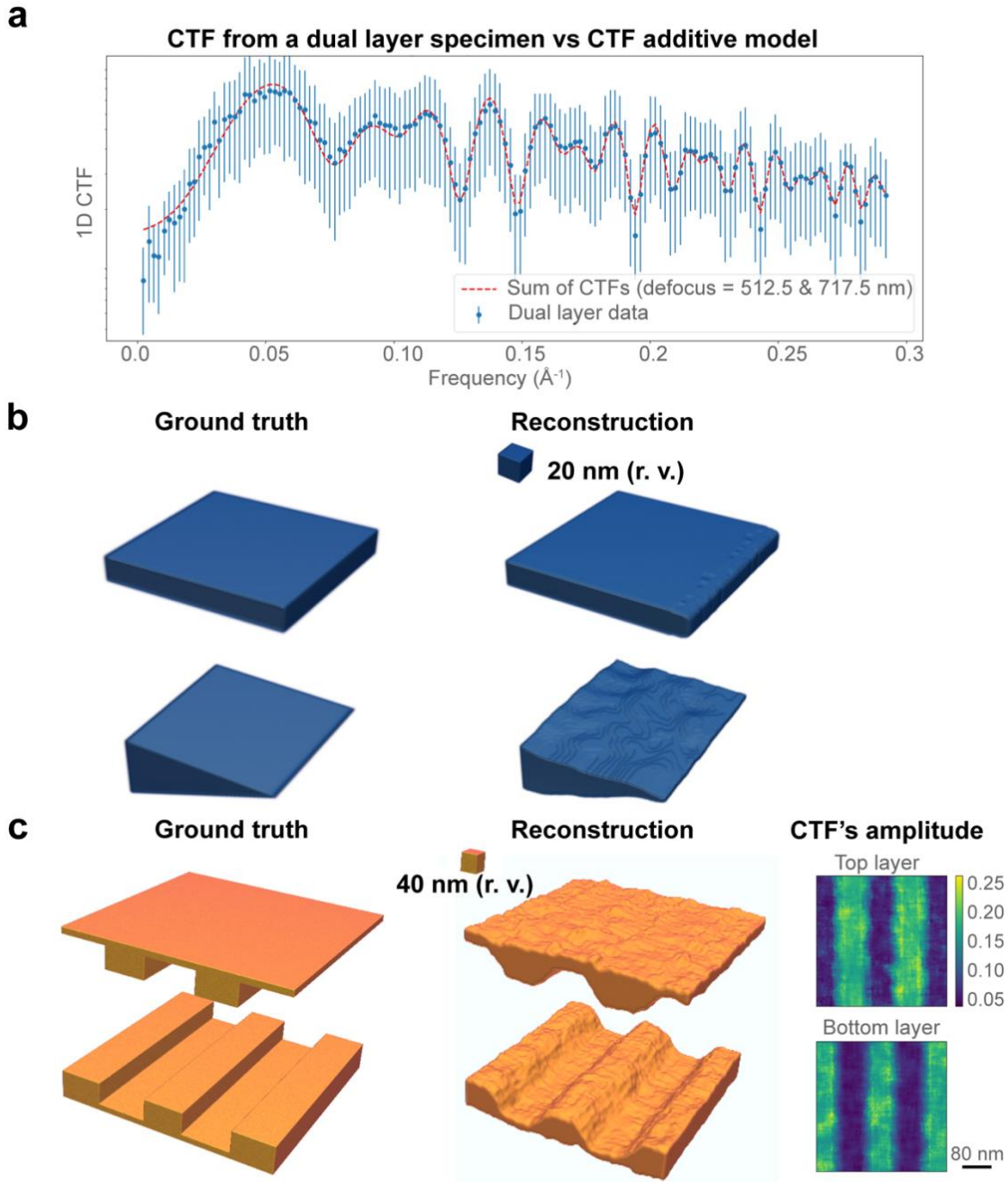

**Fig 4. Numerical validation for dual-layer pop-out metrology.** a) The radial average of Thon rings from a dual-layer specimen (simulated) and the radial average of the sum of contrast transfer functions (CTFs) from the defocus values of both layers. Both layers are 25 nm in thickness, and they are 180 nm apart; the defocus applied on the exit wave is 500 nm. Hence the centre of mass of both layers from the image plane is 512.5 nm and 717.5 nm. b) A dual-layer 3D model is simulated, and the pop-out reconstruction is generated using the prior that the thickness of the top layer is known. c) A dual-layer 3D model is simulated (total dose 20,000 e Å$^{-2}$), and the pop-out reconstruction is generated using the amplitude ratio between the additive CTFs fitted for the top and bottom layers. This reconstruction shows that the additive CTF model can also provide relative layer thickness information as well as with depth information. The resolution voxel size (r.v.) indicates the corresponding reconstruction resolution.

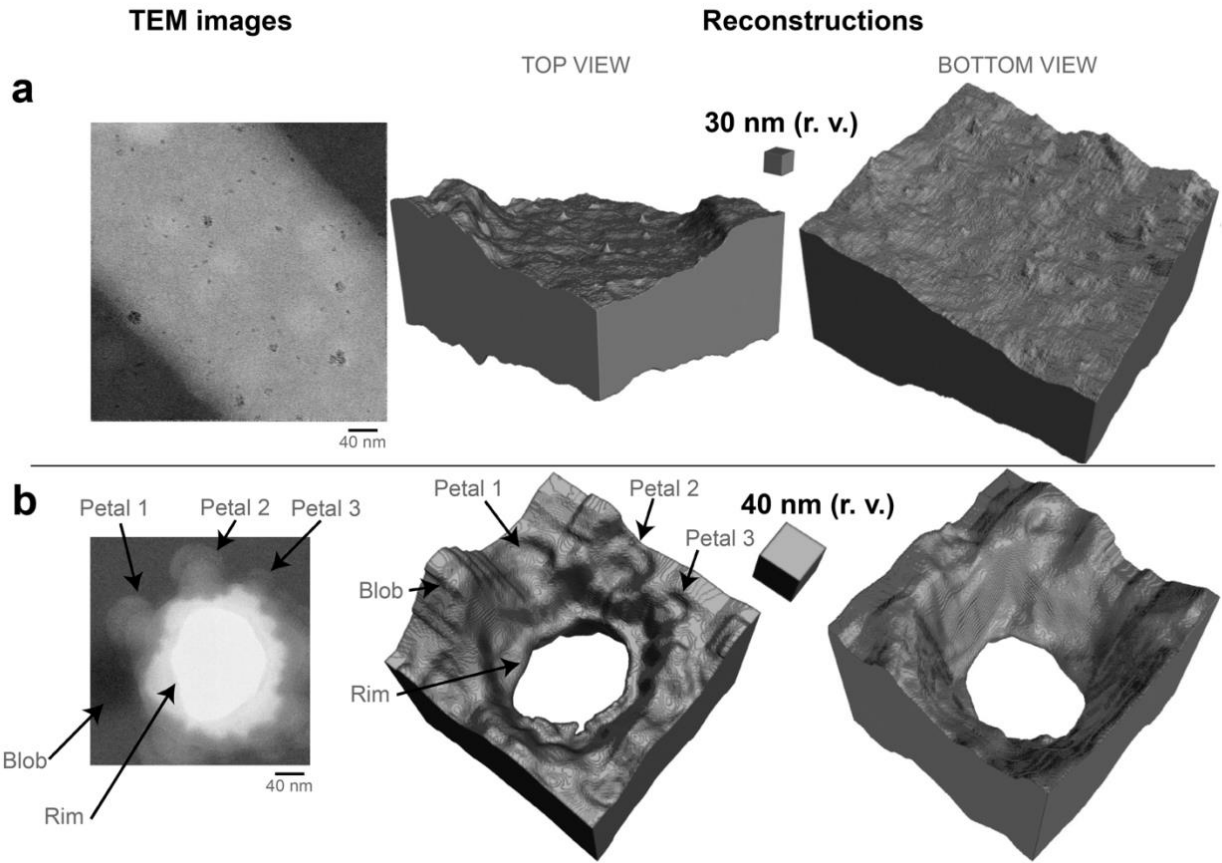

**Fig 5. Experimental validations of pop-out 3D metrology.** a) An energy-filtered bright field transmission electron microscope (BF-TEM) image (total dose 2000 e Å$^{-2}$) of a specimen with features on one side, i.e., a nano-channel etched on an amorphous SiN$_x$ membrane. The top and bottom sides of a volumetric reconstruction show that the channel is etched only on the top surface, while the bottom surface remains relatively flat. b) An energy-filtered BF-TEM image (total dose 2500 e Å$^{-2}$) of a specimen with features on either side, i.e., a nano-pit etched all the way on an amorphous SiN$_x$ membrane; the reconstruction shows that all the labelled rim, petals, and the blob of debris are present on the top surface. Although the substrate was etched from the top, the reconstruction shows that the opening of the nanopit was widened towards the bottom surface. The resolution voxel size (r.v.) indicates the corresponding reconstruction resolution.

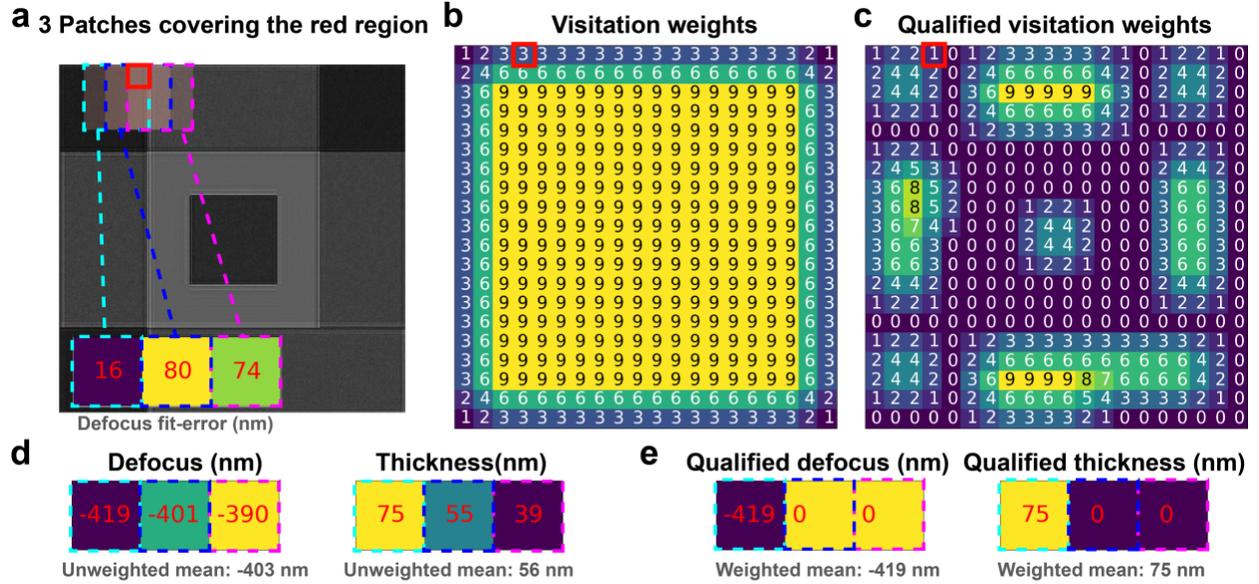

**Fig 6. Visitation weights-based running-window averaging for estimating sample thickness and depth of a local stride-length region.** a) A simulated transmission electron microscope (TEM) image showing a stride-length region of interest marked by the red square with three overlapping patches (cyan, blue, and magenta) that cover the region. The three patches correspond to the three defocus-fitting patches that apply to the red stride-length region. The inset shows the defocus fit-error values for these three patches. b) The sum of visitation weights represents the number of the corresponding patches for each of the stride-length regions. The red square here shows the same stride-length region as that in panel (a). c) The sum of qualified visitation weights after discarding the patches with fit errors larger than the lateral resolution (30 nm). The red square stride-length region only has one qualified image patch. d) The fitted defocus, and the estimated thickness values from the three patches in panel (a) that cover the red stride-length region. e) The defocus and thickness values are weighted by only the qualified patches for the red stride-length region.

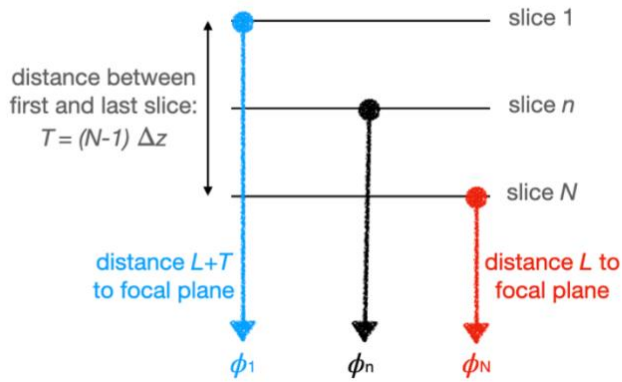 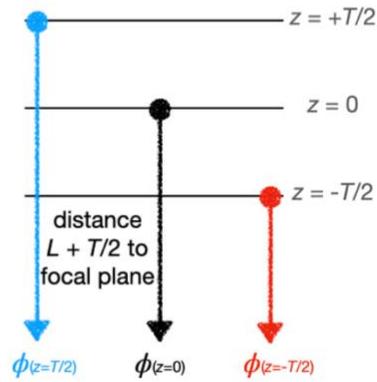

**Fig 7. Schematic illustration of the multislice formalism.** a) A slab of homogeneous scattering material of thickness $T$ is partitioned into N thin slices, each of thickness $\Delta z$. The slices scatter an incoming electron wave to produce exit wave $\phi_n$ after propagating to the rest of the specimen and further to focal place at distance $L$  b) The same set of slices represented by their position with respect to the centre of the scattering mass $z$ and produce exit wave $\phi_z$.

# Supplementary Information: Single-shot, coherent, pop-out 3D metrology

**Deepan Balakrishnan**[1,2*], **See Wee Chee**[1,3], **Zhaslan Baraissov**[1,3], **Michel Bosman**[4], **Utkur Mirsaidov**[1,2,3,4], **and N. Duane Loh**[1,2,3*]

[1]Centre for Bio-Imaging Sciences, National University of Singapore, 117557, Singapore
[2]Department of Biological Sciences, National University of Singapore, 117557, Singapore
[3]Department of Physics, National University of Singapore, 117551, Singapore
[4]Department of Materials Science and Engineering, National University of Singapore, 117575, Singapore

* Corresponding authors. email: *deepan@u.nus.edu*, *duaneloh@nus.edu.sg*

**Supplementary Note 1: Validation for effective defocus represents the centre of the scattering mass**

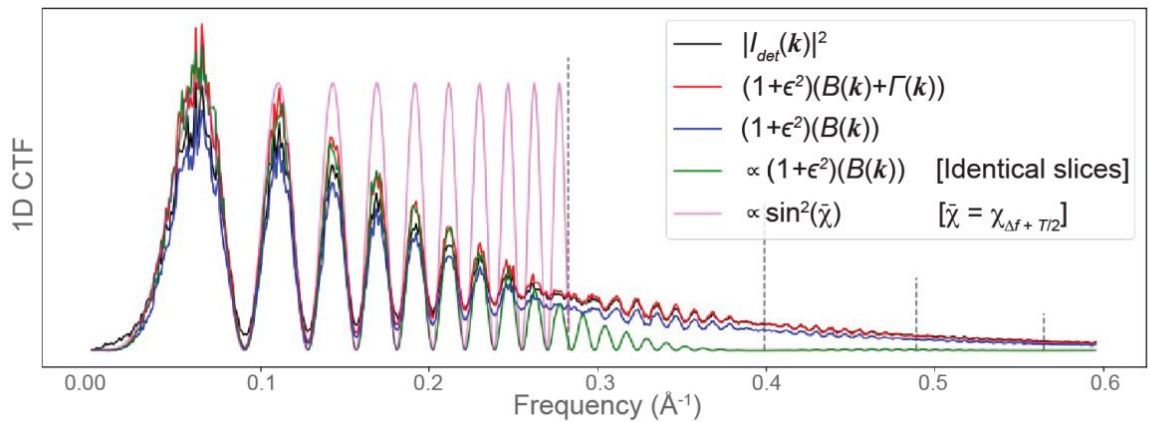

**Fig. S1 Comparison of angularly averaged power spectra from the multislice TEM micrograph (black) and from the equation (27).** The red curve from the equation (27) provides a good estimation of the power spectrum. Even when we ignore the cross-multiplication term $\Gamma(\boldsymbol{k})$ and plot the $B(\boldsymbol{k})$ term (blue), it already provides the expected nodes that occur at $\lambda k^2 T$. We can notice that if we have identical slices (green), the nodes occur at $\lambda k^2 T/2$ (dashed lines) instead of $\lambda k^2 T$. The green curve (identical slices case) is scaled down to match the amplitudes of the other curves.

# Supplementary Note 2: 2D CTF fitting with astigmatism

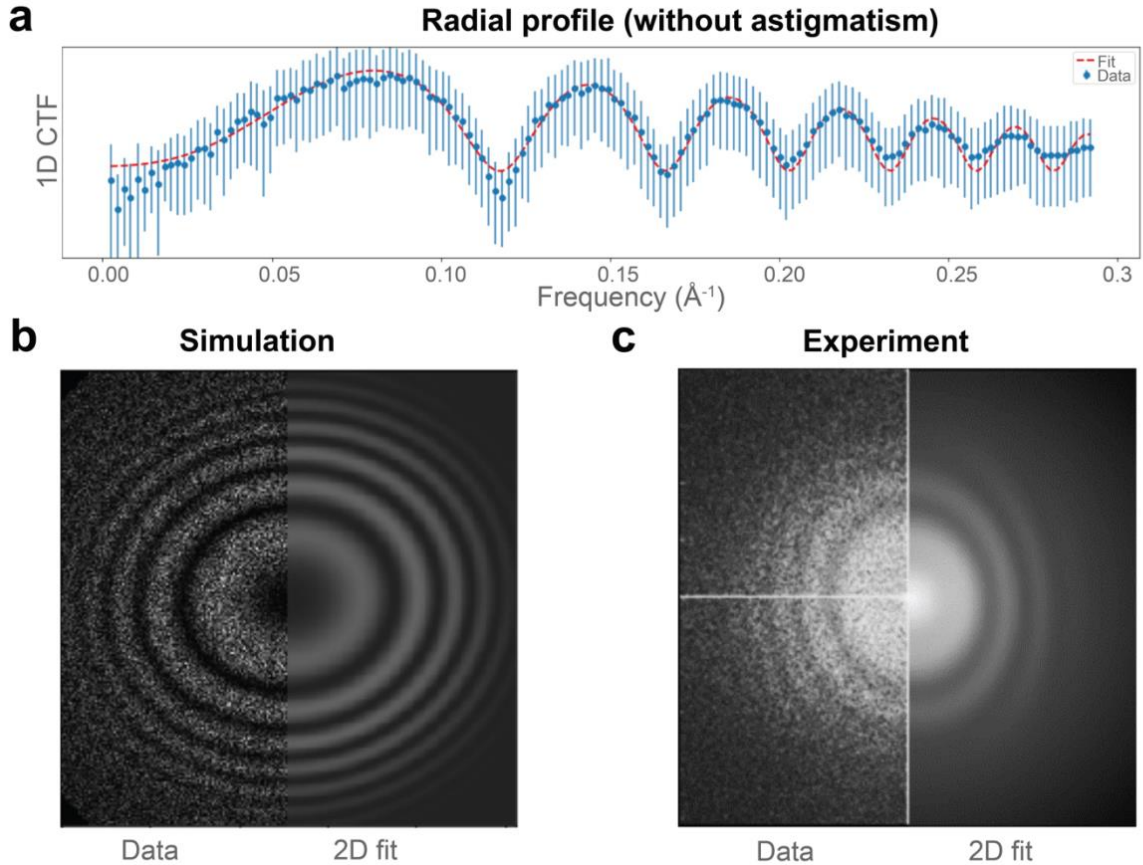

**Fig. S2 Comparison of CTFs from TEM image and their corresponding fit.** a) The radial profiles of the model fit and of the observed CTF rings from a simulated TEM image without any astigmatism; the error bars indicate the standard deviation of the image's power spectrum along each spatial resolution ring. b) A hybrid image comparing the model fitted and the observed CTF rings from a simulated TEM image with astigmatism. c) A hybrid image comparing the model fitted and the observed CTF rings from an experimental TEM image with astigmatism. The radial profile shows that the zero-crossings of the observed CTFs match those in the theoretical model, which provides validation for the curve-fitting and the model. In experiments, it is difficult to remove astigmatism completely. Hence, we introduced astigmatism in a simulated TEM image; the simulation and experimental CTF show heavy astigmatism; nevertheless, the method is robust enough to fit a theoretical CTF in both cases.

## Supplementary Note 3: Gauss envelope model vs. sinc envelope model

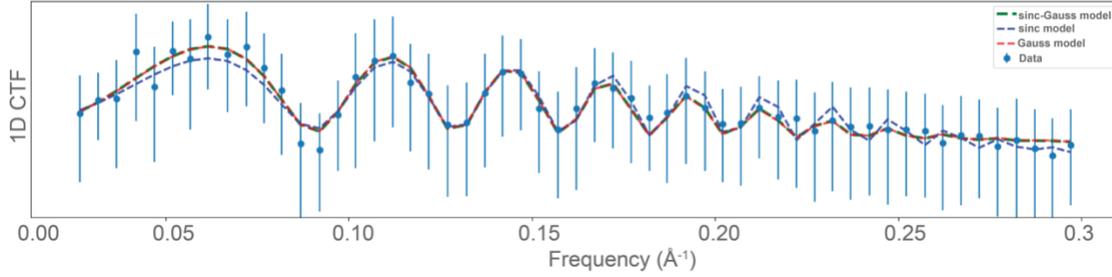

**Fig. S3 Comparison of models with a Gauss envelope and a sinc envelope functions.** The error bars indicate the standard deviation of the image's power spectrum along each spatial resolution ring.

Eq. (27) provides a theoretical sinc envelope model for fitting the CTF with the influence of specimen thickness. As we calculate the thickness information from the electron loss, we expected that a sinc model with known thickness instead of a Gaussian envelope would provide a better fit. However, many things would cause an envelope in the actual spectrum, such as spatial and temporal incoherency, a beam fall-off, a detector point spread function, sample damage, and motion blur. Hence, a generic Gaussian-only model in Eq. (4) is used to model these effects, as shown in Fig. S3. Additionally, we also compared it against a combination model comprising both sinc and Gauss envelopes, as well as the Gaussian-only model. Ultimately, we chose the Gaussian model for simplicity and fit efficiency.

## Supplementary Note 4: Experimental results of TEM images without energy filter

The thickness of the specimen is calculated from the electrons absorbed by the specimen. However, imaging without an energy filter would let all the electrons, including the inelastically scattered electrons, reach the detector. These inelastically scattered electrons form a background, which does not contribute to the image contrast. Hence, both the calibrated electron inelastic mean free path $\ell_{\text{mfp}}$ and the intensity at the detector $I_{t(x,y)}$ would be erroneous, and we cannot rely on the thickness map determined as shown in Eq. (S1).

$$T'_{(x,y)} = \ell'_{\text{mcp}} ln(I_0/I'_{t(x,y)}) \tag{S1}$$

An energy-filtered TEM image would resolve this issue as the inelastically scattered electrons are removed.

### 4.1. Amorphous silicon nitride nanochannel and nanopit

The TEM images of an amorphous $SiN_x$ nanochannel and a nanopit are images without energy filter and reconstructed with pop-out 3D metrology (Fig. S4). The reconstructions show that the defocus determination is robust enough to pop-out the shape of the channel and the pit. Though the reconstruction is not quantitatively valid as the pit and channel's depth are inaccurate, the channel and the pit are popped-out only on the substrates' etched side and not visible on the bottom side of the substrates. The edge artefacts visible on the bottom side of the substrates are within the reconstruction voxel size. However, due to the limitation that we cannot resolve whether there is any material in the centre of the pit or not, using pop-out without an energy filter produces erroneous reconstructions in such cases.

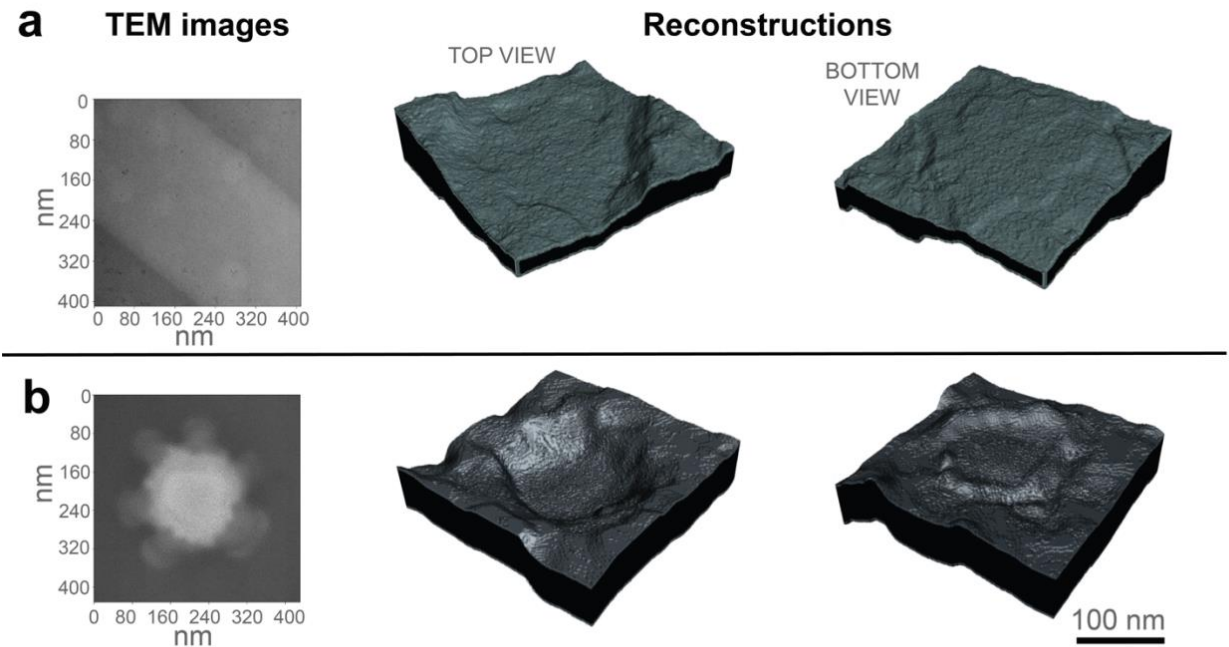

**Fig. S4 Experimental demonstration of the pop-out 3D metrology on TEM images without an energy filter.** a) TEM image of an amorphous $SiN_x$ (a-$SiN_x$) nanochannel (total dose of 283 e Å$^{-2}$), and the volume rendering of the 3D structure. b) TEM image of an amorphous $SiN_x$ nanopit (total dose of 850 e Å$^{-2}$), and the volume rendering of the 3D structure. The flat bottom of the rendered 3D structure nanochannel validates the depth estimation so the proposed method can reconstruct the 3D structures qualitatively, even from a single BF-TEM image without energy filtering. However, due to the presence of inelastically scattered electrons in the centre of the nano-pit TEM image, the pop-out could not resolve whether there is any material in the pit or not.

### 4.2. Polycrystalline pillars on an amorphous silicon nitride substrate

The polycrystalline material comparatively produces either a few Bragg peaks or diffraction rings in the Fourier space. Hence the CTF rings should be fitted for the resolution range lesser than the diffraction ring's resolution, and the SNR of the CTF rings is also very low. Because of this reduced fit range, polycrystalline materials require a higher electron dose than amorphous material, and the diffraction rings hinder the CTF undulation, which limits the resolution. The high-resolution TEM (HRTEM) image of a couple of nanopillars of height and diameter of 50 nm, the thickness map, the defocus map, and the volume rendering are shown in Fig. S5. The power spectrum from the amorphous substrate shows prominent Thon rings. However, the polycrystalline pillar's power spectrum shows a halo diffraction ring with more prominent Bragg peaks. The Thon rings are visible only at the lower frequencies as most of the electrons are sampled near the Bragg peaks. In the CTF fitting process, only the lower frequencies within the diffraction ring are used.

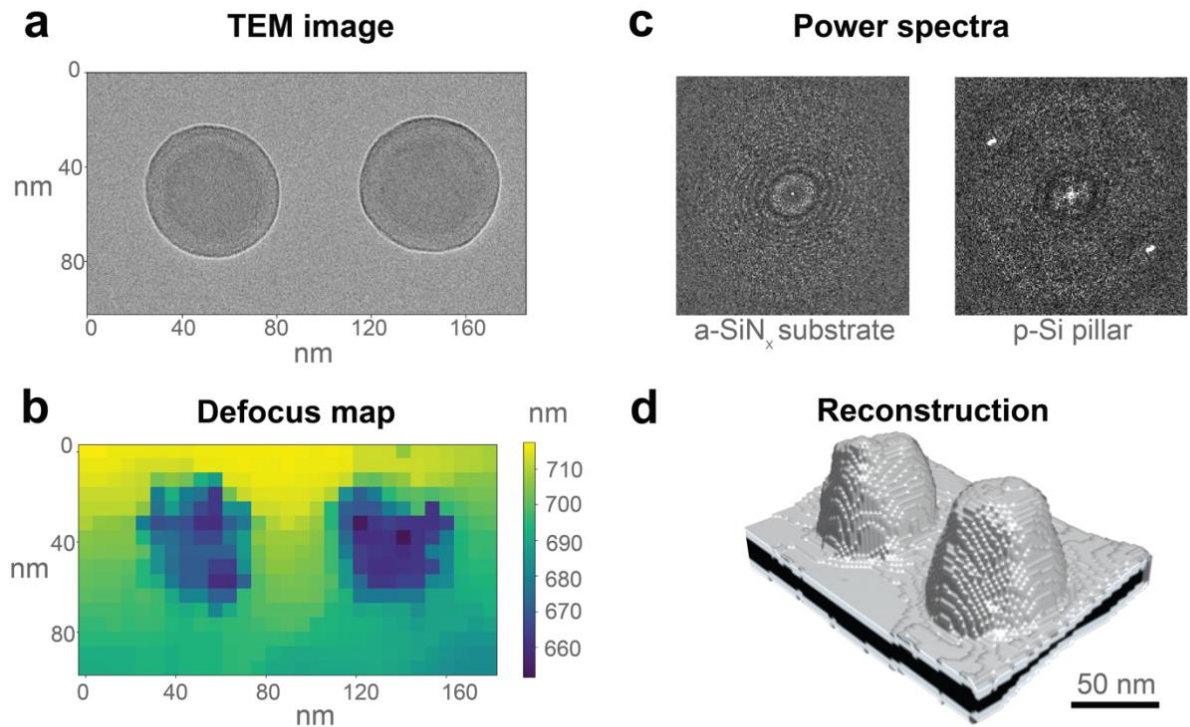

**Fig. S5 Experimental demonstration of pop-out 3D metrology on polycrystalline material.** a) HRTEM image of a polycrystalline Si nanopillar on an a-SiN$_x$ membrane (total dose of 3200 e Å$^{-2}$). b) Observed CTFs from the amorphous substrate and polycrystalline pillars. c) The defocus map, which is determined with astigmatism. d) The volume rendering of the reconstructed 3D structure.

The amorphous SiN$_x$ provided a smooth, noiseless defocus map, but the polycrystalline pillars' defocus values are noisy due to the aforementioned reasons. We can notice a small tilt in the substrate from the defocus values, which cannot be picked from the TEM image of the thickness map. The reconstruction is not smooth, and the curvature of pillars is pixelated due to poor lateral resolution as the polycrystalline Si requires a larger patch to sample frequencies within the Bragg peak resolution. Although pop-out metrology can be applicable to polycrystalline materials, the current implementation's resolution is limited by the diffraction contrast.

## Supplementary Note 5: Nano-pit simulation to validate the experiment reconstruction

The nanopit reconstruction shown in the manuscript (Fig. 5b) reveals that the nanopit has a double-cone structure. We have simulated a TEM image of a perfect cylindrical nanopit (hole) with a sharp edge and reconstructed it with the pop-out 3D metrology. Both microscope parameters and specimen properties are matched with the experiment data, and the reconstruction is carried out with the same patch size (40 nm). Fig. S6a shows that the reconstruction's chamfered edge artefact (40 nm) is within the reconstructed voxel size. However, the experiment reconstruction in Fig. S6b shows that the slope size (90 nm) is much larger than the reconstructed voxel size. Hence, we can validate that the double-cone shape is not an artefact of our algorithm, and is an actual feature of the nanopit.

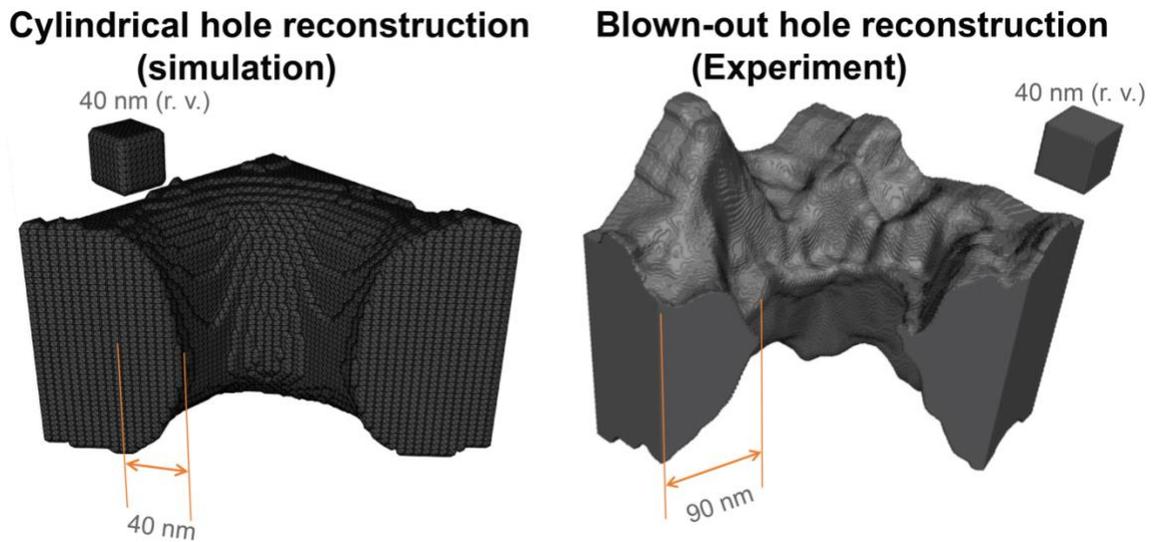

**Fig. S6 Validation for the double-cone structure resolved by pop-out in a nanopit reconstruction.** a) The reconstruction of the simulated nanopit shows that the size of the chamfered edge artefact (40 nm) from the filtering process is same as the reconstruction voxel size (40 nm). b) The reconstruction of the experimental nanopit in Fig. 5b shows that the size of the slope (90 nm) is larger than the reconstruction voxel size (40 nm).

## Supplementary Note 6: Challenges with complex 3D structures

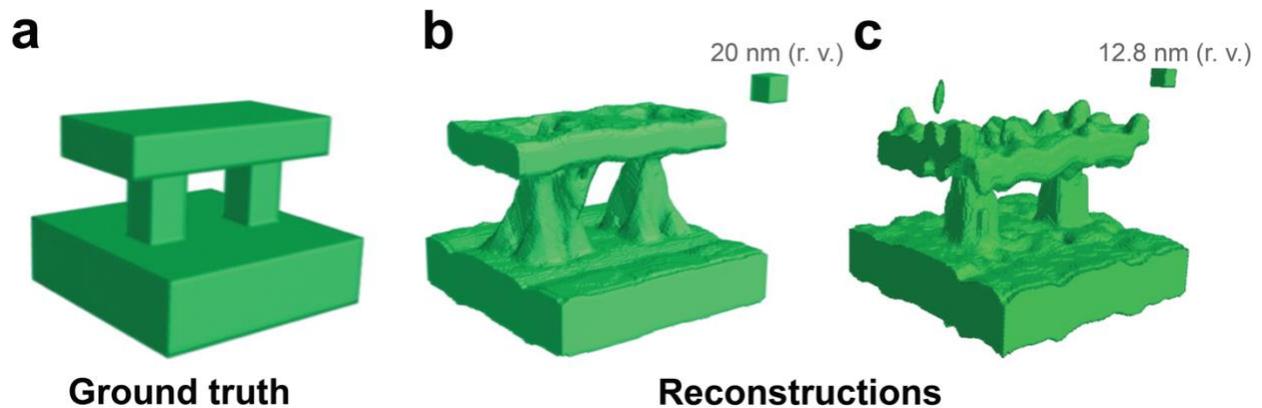

**Fig. S7 A connected dual-layer specimen reconstruction to demonstrate the challenges in dual-layer pop-out metrology.** a) The ground truth of the specimen. b) A dual-layer reconstruction with a 20 nm patch size, and c) a dual-layer reconstruction with a 12.8 nm patch size.

As mentioned in the main text, a CTF-fitting patch with more than one defocus (as in a step structure where the patch partly covers two steps) gives an erroneous fit value. This issue can be resolved for a single-layer pop-out by identifying the erroneous fit from the error map and interpolating the defocus values from the neighbourhood region. However, for a complex specimen, as shown above, the pillars are a single layer that should be fitted with a single CTF model, and the regions around the pillars are dual-layered, which requires the additive CTF model to fit defocus from both layers. We fitted the entire TEM image with both the single CTF model and additive CTF model, and then the error values in the fit were used to pick the single-layered and dual-layered regions. This approach worked for the membrane regions except for the region around the pillars (Fig. S7b). The error values in either model are so high in these regions because the fitting patch partly covers the single-layer thick pillar and partly covers the dual-layer membranes. Reducing the patch size would confine this problem to a small region around the pillar. However, the smaller patch size could not sample frequencies fine enough to fit two CTFs accurately. The thin top membrane has poor reconstruction in Fig. S7c due to the inaccurate dual-layer fit caused by a smaller patch size (12.8 nm).